\documentclass[10pt]{iopart}

\usepackage{citesort}
\usepackage{graphicx}
\usepackage{algorithm}
\usepackage{algpseudocode}
\usepackage{algpascal}
\usepackage{color}

\definecolor{azu}{rgb}{0.0,0.4,1.0}

\begin{document}

\title[Micromagnetics of rare-earth efficient permanent magnets]{Micromagnetics of rare-earth efficient permanent magnets}

\author{Johann Fischbacher$^1$, Alexander Kovacs$^1$, Markus Gusenbauer$^1$, Harald Oezelt$^1$, Lukas Exl$^{2,3}$, Simon Bance$^4$, Thomas Schrefl$^1$}

\address{$^1$ Department for Integrated Sensor Systems, Danube University Krems, Viktor Kaplan Stra\ss e 2, 2700 Wiener Neustadt, Austria}
\address{$^2$ Faculty of Mathematics, University of Vienna, Oskar-Morgenstern-Platz 1, 1090 Wien, Austria}
\address{$^3$ Institute for Analysis and Scientific Computing, Vienna University of Technology, Wiedner Hauptstra\ss e 8-10, 1040, Wien, Austria}
\address{$^4$ Seagate Technology, 1 Disc Drive, Springtown, Derry, BT48 0BF Northern Ireland, UK}
\ead{thomas.schrefl@donau-uni.ac.at}
\vspace{10pt}
\begin{indented}
\item[]January 2018
\end{indented}

\begin{abstract}
The development of permanent magnets containing less or no rare-earth elements is linked to profound knowledge {of} the coercivity mechanism. Prerequisites for a promising permanent magnet material are a high spontaneous magnetization and a sufficiently high magnetic anisotropy. In addition to the intrinsic magnetic properties the microstructure of the magnet plays a significant role {in establishing} coercivity. The influence of the microstructure on coercivity, remanence, and energy density product can be understood by {using} micromagnetic simulations. With advances in computer hardware and numerical methods, hysteresis curves of magnets can be computed {quickly} so that the simulations can {readily} provide {guidance} for the development of permanent magnets. {The potential of rare-earth reduced and free permanent magnets is investigated using micromagnetic simulations. The results show excellent hard magnetic properties can be achieved in grain boundary engineered NdFeB, rare-earth magnets with a ThMn$_{12}$ structure,  Co-based nano-wires, and L1$_0$-FeNi provided that the magnet's microstructure is optimized.}
\end{abstract}

%
% Uncomment for keywords
\vspace{2pc}
\noindent{\it Keywords}: micromagnetics, permanent magnets, rare earth
%
% Uncomment for Submitted to journal title message
%\submitto{\JPD}
%
% Uncomment if a separate title page is required
\maketitle
% 
% For two-column output uncomment the next line and choose [10pt] rather than [12pt] in the \documentclass declaration
\ioptwocol

\section{Introduction}

\subsection{Rare-earth reduced permanent magnets}

\begin{figure}
\includegraphics[width=8.0cm]{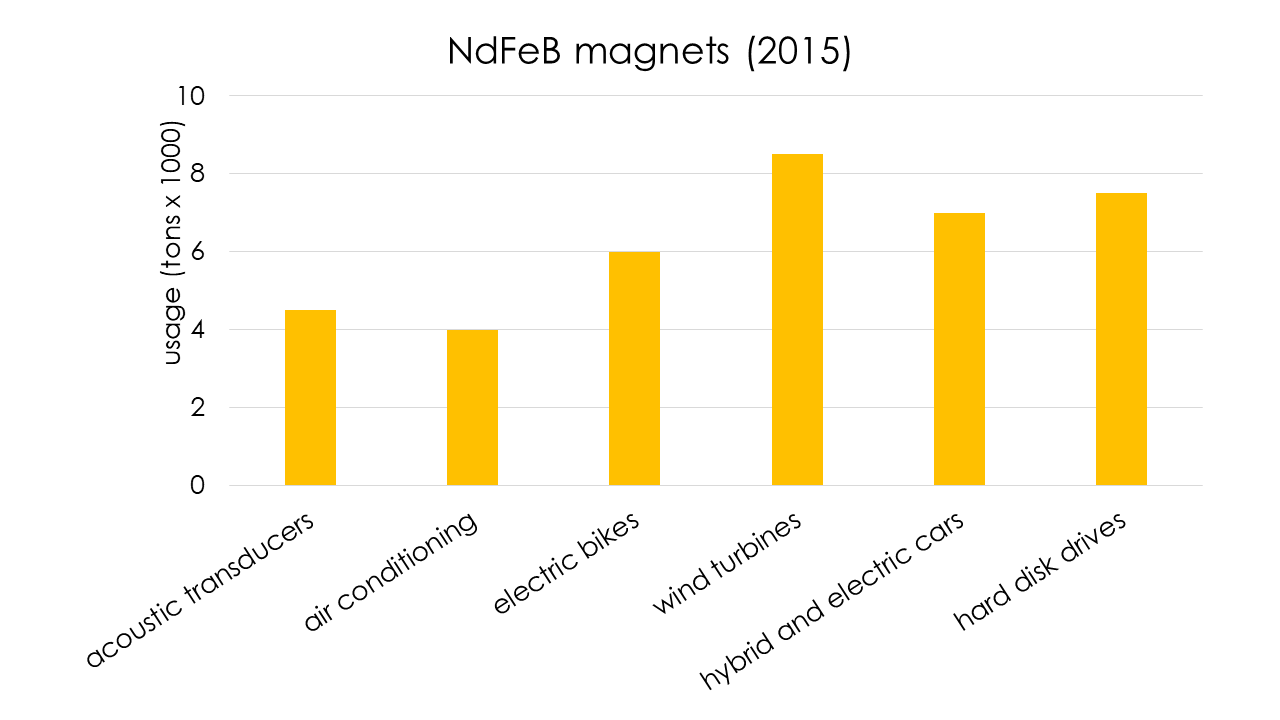}
\caption{\label{fig_use2015} Usage of NdFeB magnets in the six major markets in the year 2015. Data taken from \cite{Constantinides2016}.}
\end{figure}

High performance permanent magnets are a key technology for modern society. High performance magnets are distinguished by (i) the high magnetic field they can create and (ii) their high resistance to opposing magnetic fields. A prerequisite for these two characteristics are proper intrinsic properties of the magnet material: A high spontaneous magnetization and high magneto-crystalline anisotropy. The intermetallic phase Nd$_2$Fe$_{14}$B \cite{sagawa1984new,croat1984pr} fulfills these properties. Today NdFeB-based magnets dominate the high performance magnet market. {In the following we will use ``Nd$_2$Fe$_{14}$B'' when we refer to the intermetallic phase and ``NdFeB'' when we refer to a magnet which is based on Nd$_2$Fe$_{14}$ but contains additional elements.}  There are six major sectors which heavily rely on rare-earth permanent magnets \cite{Constantinides2016}. The usage of NdFeB is summarized in figure \ref{fig_use2015} based on data given by Constantinides \cite{Constantinides2016} for 2015. Modern acoustic transducers use NdFeB magnets. Speakers are used in cell phones, consumer electronic devices, and cars. The total number of cell phones {that are} shipped per year is reaching 2 {billion}. Air conditioning is a growing market. Around 100 million units are shipped every year. Each unit uses about three motors with NdFeB magnets. NdFeB magnets are essential to sustainable energy production and eco-friendly transport. The generator of a direct drive wind mill requires high performance magnets of 400 kg/MW power; and on average a hybrid and electric vehicle needs 1.25 kg of high end permanent magnets \cite{yang2017ree}. Another rapidly growing market is electric bikes with 33 million global sales in 2016. For {a} long time NdFeB magnets have been used in hard disk drives. Hard disk drives use bonded NdFeB magnets in the motor that spins the disk and sintered NdFeB magnets for the voice coil motor that moves the arm. There are around 400 million hard disk drives shipped every year.    

\begin{figure}
\includegraphics[width=8.0cm]{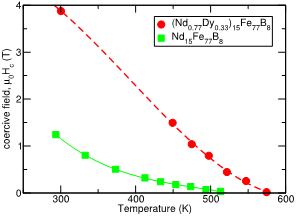}
\caption{\label{fig_HcDy} Coercive field of Nd$_{15}$Fe$_{77}$B$_8$ and (Nd$_{0.77}$Dy$_{0.33}$)$_{15}$Fe$_{77}$B$_8$ magnets as function of temperature. Data taken from \cite{sagawa1987dependence}.}
\end{figure}

In many applications the NdFeB magnet are used at elevated temperature. For example, the operating temperature of the magnet in the motor/generator block {of hybrid vehicles} is at about 450 K. Though Nd$_2$Fe$_{14}$B {($T_\mathrm{c} = 558$~K)} shows excellent properties at room temperature its Curie temperature {$T_\mathrm{c}$} is much lower than those of SmCo$_5$ {($T_\mathrm{c} = 1020$~K)} or Sm$_2$Co$_{17}$ {($T_\mathrm{c} = 1190$~K)} magnets \cite{coey_2010}. Therefore the anisotropy field and the coercive field of Nd$_2$Fe$_{14}$B rapidly decays with {increasing} temperature. In order to compensate this loss, some of the magnet's Nd is replaced with heavy rare earths such as Dy. Figure \ref{fig_HcDy} compares the  coercive field of conventional Dy-free and Dy-containing NdFeB magnets {as a function of temperature}. (NdDy)FeB magnets, containing {around} 10 weight percent Dy, can reach coercive fields $\mu_0 H_{\rm c} > 1$~T at 450~K. However, since the rare-earth crisis \cite{sprecher2017novel} the rare-earth prices {have become more} volatile. During 2010 and 2011 the Dy price peaked and increased by a factor of 20 \cite{sander2017}. Only four percent of the primary rare-earth production comes from outside China \cite{sprecher2017novel}. Because of supply risk and increasing demand, Nd and Dy are considered to be critical elements \cite{king2016rare}. In order to cope with the supply risk, magnet producers and users aim for rare-earth free permanent magnets. With respect to the magnet's performance, rare-earth free permanent magnets may fill a gap between ferrites and NdFeB magnets \cite{coey2012permanent}. An alternative goal {is} magnets with less rare earth than (NdDy)FeB magnets but comparable magnetic properties \cite{nakamura2017current}. 

Possible routes to achieve these goals are:
\begin{itemize}
\item 
Shape anisotropy based permanent magnets;
\item
Grain boundary diffusion; 
\item 
Improved grain boundary phases;
\item
Nanocomposite magnets;
\item
Alternative hard magnetic compounds.
\end{itemize}
In this work we will use micromagnetic simulations, in order to address various design issues for rare-earth efficient permanent magnets. Micromagnetic simulations are an important tool to understand coercivity mechanisms in permanent magnets. With the advance of hardware for parallel computing \cite{exl2014labonte,sepehri2016micromagnetic,tsukahara2016large} and the improvement of numerical methods \cite{fischbacher2017nonlinear,tanaka2017speeding,erokhin2017optimization}, micromagnetic simulations can take into account the microstructure of the magnet and thus help to understand how the interplay between intrinsic magnetic properties and microstructure impacts coercivity.  

\subsection{Key properties of permanent magnets}

The primary goal of a permanent magnet is to create a magnetic field in the air gap of a magnetic circuit. The energy stored in the field outside of a permanent magnet can be related to its magnetization and to its shape. According to Maxwell's equations the magnetic induction $\mathbf B$ is divergence-free (solenoidal): $\nabla \cdot \mathbf B = 0$ and in the absence of any current the magnetic field $\mathbf H$ is curl-free (irrotational): $\nabla \times \mathbf H = 0$. The volume integral of the product of a solenoidal and irrotational vector field over all space is zero, when the corresponding vector and scalar potentials are regular at infinity \cite{brown1962magnetostatic}. This is the case when 
\begin{equation}
\label{eq_BmuMplusH}
\mathbf B = \mu_0 (\mathbf M + \mathbf H)
\end{equation}
is the magnetic induction due to the magnetization $\mathbf M$ of a magnet. Here $\mu_0 = 4\pi \times 10^{-7}$~Tm/A is the permeability of vacuum. The magnetostatic energy in a volume $V_\mathrm{a}$ of free space, where $\mathbf M = 0$ and $\mathbf B = \mu_0 \mathbf H$, is $E_{\mathrm {mag,a}} = (\mu_0/2)\int_{V_\mathrm{a}} \mathbf H^2 dV$. Splitting the space into the volume inside the magnet, $V_{\mathrm i}$, and $V_{\mathrm a}$, we have $\int \mathbf B \cdot \mathbf H dV = \int_{V_\mathrm{a}} \mu_0 \mathbf H^2 dV + \int_{V_\mathrm{i}} \mathbf B \cdot \mathbf H dV= 0$ or
\begin{equation}
\label{eq_BHoutside}
E_{\mathrm {mag,a}} = -\frac{1}{2} \int_{V_\mathrm{i}} \mathbf B \cdot \mathbf H dV. 
\end{equation} 
Since the left-hand side of equation (\ref{eq_BHoutside}) is positive, $\mathbf B$ and $\mathbf H$ must point in opposite directions inside the magnet. Approximating the magnetic induction $\mathbf B$ and the magnetic field $\mathbf H$ by a uniform vector field inside the magnet, we can write $E_{\mathrm {mag,a}} = (1/2)\int_{V_\mathrm{i}} (BH) dV$, where $B=\left|\mathbf B\right|$ and $H=\left| \mathbf H \right| $. We see that we can increase the energy stored in its external field either by increasing the magnet's volume $V_{\mathrm i}$ or by increasing the product $(BH)$, which is referred to as energy density product \cite{fidler1997review}. It is defined as the product of the magnetic induction $B$ and the corresponding opposing magnetic field $H$ \cite{buschow2003physics} and is given in units of J/m$^3$. When there are no field generating currents, the magnetic field inside the magnet 
\begin{equation}
\label{eq_demagfield}
\mathbf H = - N \mathbf M
\end{equation}
depends on the magnet's shape which can be expressed by the demagnetizing factor $N$. We further assume that the magnet is saturated and there are no secondary phases so that $ \left| \mathbf M \right| = M_{\mathrm s}$, where $M_{\mathrm s}$ is the spontaneous magnetization of the material. Using equations (\ref{eq_BmuMplusH}) and (\ref{eq_demagfield}) we express the energy density product as $(BH) = \left| \mu_0 (\mathbf M - N \mathbf M) \right| \left|- N \mathbf M \right|  = \mu_0 (1 - N)N M_\mathrm{s}^2$ \cite{coey2012permanent,skomski2016magnetic}. When maximized with respect to $N$ this gives the maximum energy density product of a given material
\begin{equation}
\label{eq_bhmax}
(BH)_\mathrm{max} = \frac{1}{4} \mu_0 M_\mathrm{s}^2
\end{equation}
for $N=1/2$. It is worth to check the shape of a magnet with a demagnetizing factor of $1/2$. Let us assume a magnet in form of a prism with dimensions  $l \times l \times pl$ which is magnetized along the edge with length $pl$. Then a simple approximate equation for the demagnetizing factor is $N = 1/(2p+1)$ \cite{sato1989simple}. Therefore, the optimum shape of a magnet that results in the maximum energy density product is a flat prism with dimensions $l \times l \times 0.5l$, which is twice as wide as high. Many modern magnets have this shape.

\begin{figure}
\includegraphics[width=8.0cm]{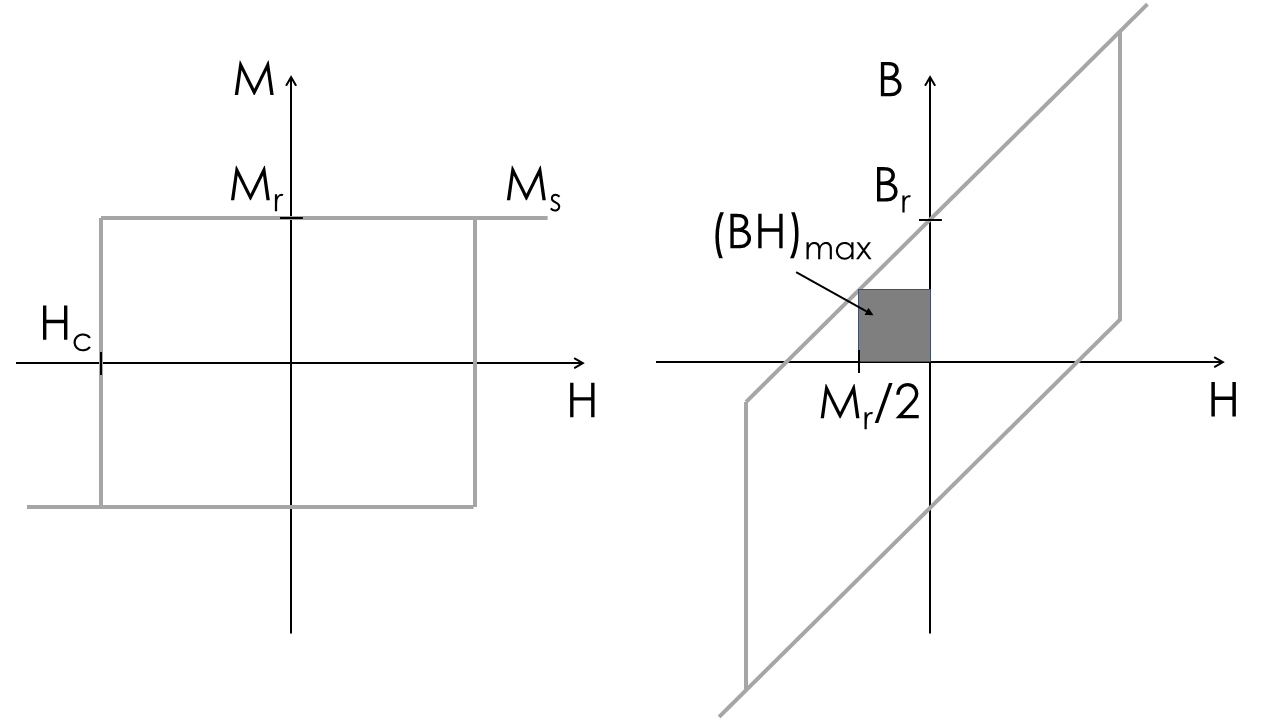}
\caption{\label{fig_bhmax}The maximum energy density product $(BH)_\mathrm{max}$ is given by the area of the largest rectangle that fits below the 2nd quadrant of the $B(H)$ curve. Left: $M(H)$ loop, right: $B(H)$ loop of an ideal magnet.}
\end{figure}

When there is no drop of the magnetization with {increasing} opposing field until $H > M_\mathrm{s}/2$, the energy density product reaches its maximum value given by equation (\ref{eq_bhmax}). 
In this case, the magnetic induction $B$ as function of field is a straight line. For an ideal loop as shown in figure \ref{fig_bhmax} the remanent magnetization, $M_\mathrm{r}$, equals the spontaneous magnetization, $M_\mathrm{s}$. In some materials, magnetization reversal may occur at fields lower than half the remanence. When $H_{\mathrm c} < M_\mathrm{r}/2$, the energy density product is limited by the coercive field, $H_{\mathrm c}$. Similarly, the maximum value for $(BH)_\mathrm{max}$ is not reached, when $M(H)$ is not square but decreases with increasing field $H$. 

A higher energy density product reduces volume and weight of the permanent-magnet-containing device making it an important figure of merit. Other decisive properties are the remanence, the coercive field, and the loop squareness.

\subsection{Permanent magnets and intrinsic magnetic properties}

A magnetic material suitable for a permanent magnet must have certain intrinsic magnetic properties. From inspection of figure \ref{fig_bhmax} we see that a good permanent magnet material requires a high spontaneous magnetization $M_\mathrm{s}$ and a uniaxial anisotropy constant $K$ that creates a coercive field
\begin{equation} 
\label{eq_Hc}
H_\mathrm{c} > \frac{M_\mathrm{s}}{2}.
\end{equation} 
The theoretical maximum for the coercive field is the nucleation field \cite{brown1957criterion}
\begin{equation}
\label{eq_Hnuc}
H_{\mathrm N} = \frac{2 K}{\mu_0 M_\mathrm{s}}
\end{equation}
for magnetization reversal by uniform rotation of a small sphere. Equations (\ref{eq_Hc}) and (\ref{eq_Hnuc}) give the condition $K > \mu_0 M_\mathrm{s}^2/4$. In other words, the anisotropy energy density, $K$, should be larger than the maximum energy density product, $(BH)_\mathrm{max}$. 
For most magnetic materials this condition is not sufficient \cite{coey2012permanent}. There are two stronger conditions for the magnetocrystalline anisotropy constant.

To be able to make a permanent magnet or its constituents in any shape, the nucleation field must be higher than the maximum possible demagnetizing field. The demagnetizing factor of a thin magnet {approaches} 1 and the magnitude of the demagnetizing field approaches $M_\mathrm{s}$ which gives the condition $K > \mu_0 M_\mathrm{s}^2/2$. This is often expressed in terms of the quality factor $Q = {2 K}/({\mu_0 M_\mathrm{s}^2})$, which was introduced in the context of bubble domains in thin films \cite{thiele1970theory,slonczewski1973statics}. For $Q > 1$ stable domains are formed and the magnetization points either up or down along the anisotropy axis perpendicular to the film plane. Otherwise, the demagnetizing field would cause the magnetization to lie in plane. 

\begin{figure}
\includegraphics[width=8.0cm]{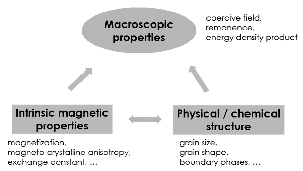}
\caption{\label{fig_microstructure}Both the intrinsic magnetic properties and the physical/chemical structure of the magnet determine coercive field, remanence and energy density product.}
\end{figure}

The maximum possible coercive field is never reached experimentally. This phenomenon is usually referred to as Brown's paradox \cite{brown1945virtues,shtrikman1960resolution}. Imperfections are one reason for the reduction of the coercive field with respect to its ideal value. In the presence of defects with zero magnetocrystalline anisotropy, the coercive field may reduce to $H_{\mathrm c} = H_{\mathrm N}/4$. Plugging this limit for the coercive field into equation (\ref{eq_Hc}) gives the condition $K > \mu_0 M_\mathrm{s}^2$ for the anisotropy constant. This corresponds to the empirical law $\kappa > 1$ for many hard magnetic phases of common permanent magnets \cite{coey2012permanent}, where $\kappa = \sqrt{{K}/({\mu_0 M_\mathrm{s}^2})}$ is the hardness parameter \cite{coey1995new}.

The key figures of merit of permanent magnets such as the coercive field, the remanence, and the energy density product are extrinsic properties. They follow from the interplay of intrinsic magnetic properties and the granular structure of the magnet which is schematically shown in figure \ref{fig_microstructure}. Thus, in addition to the spontaneous  magnetization $M_\mathrm{s}$, magnetocrystalline anisotropy constant $K$, and the exchange constant $A$, a well-defined physical and chemical structure of the magnet is essential for excellent permanent-magnet properties. 

\begin{figure}
\includegraphics[width=8.0cm]{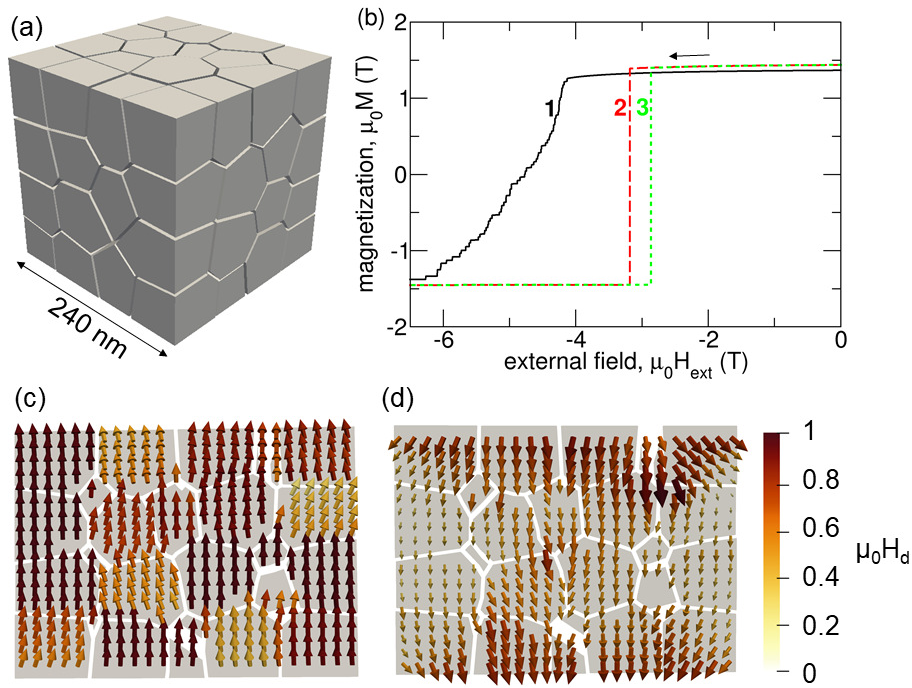}
\caption{\label{fig_microEffects}(a) Microstructure used for the simulation of microstructural effects that decrease the coercive field. (b) \emph{1~misalignment (solid line) :}  non-magnetic grain boundary phase, demagnetizing field switched off; \emph{2~Defects(dashed~line):} Weakly ferromagnetic grain boundary phase and demagnetization field switched off; \emph{3~Demagnetizing effects:} Weakly ferromagnetic grain boundary phase and demagnetization field switched on. The vector fields in (c) and (d) show magnetization and the demagnetizing field before switching for case 3.}
\end{figure}

Empirically, the effects that reduce the coercive field with respect to the ideal nucleation field are often written as \cite{kronmuller2003micromagnetism,fischbacher2017limits} 
\begin{equation}
\label{eq_kroni}
H_\mathrm{c} = \alpha_K \alpha_\psi H_\mathrm{N} - N_\mathrm{eff} M_\mathrm{s} - H_\mathrm{f}.
\end{equation} 
The coefficients $\alpha_K$ and $\alpha_\psi$ express the reduction in coercivity due to defects and misorientation, respectively \cite{kronmuller1988analysis}.  The microstructural parameter $N_\mathrm{eff}$ is related to the effect of the local demagnetization field near sharp edges and corners of the microstructure \cite{gronefeld1989calculation}. The fluctuation field $H_\mathrm{f}$ gives the reduction of the coercive field by thermal fluctuations \cite{givord1988coercivity}. 

Let us look at an example. Figure \ref{fig_microEffects}a shows the microstructure of a nanocrystalline Nd$_2$Fe$_{14}$B magnet used for micromagnetic simulations to identify the different effects that reduce coercivity. For the set of material parameters used ($K = 4.3$~MJ/m$^3$, $\mu_0 M_\mathrm{s}=1.61$~T, $A = 7.7$~pJ/m), the ideal nucleation field is $\mu_0 H_\mathrm{N} = 6.7$~T. The 64 grains were generated from a centroid Voronoi tessellation \cite{quey2011large}. The average grain size was 60~nm. The anisotropy directions were randomly distributed within a cone with an opening angle of 15 degrees. The grain boundary phase of NdFeB magnets contains Fe and is weakly ferromagnetic \cite{murakami2014magnetism,zickler2015nanoanalytical,zickler2017combined}. In addition to magnetostatic interactions between the grains, the grains are also weakly exchange coupled. In our simulations the grain boundary phase was 3 nm thick. The magnetocrystalline anisotropy constant $K$ of the grain boundary phase was zero. Its magnetization and exchange constant were $\mu_0 M_\mathrm{s}=0.5$~T and $A = 7.7$~pJ/m for cases 2 to 4.  

In numerical micromagnetics we can artificially switch physical effects on or off and thus gain a deeper understanding {of} how the various effects impact magnetization reversal. We start with the granular system whereby the grains are separated by a nonmagnetic grain boundary phase and the magnetostatic energy term is switched off. Thus, there are no demagnetizing fields and no magnetostatic interactions. The grains are isolated and there are no defects. The solid line (case 1) of figure \ref{fig_microEffects}b shows the influence of misalignment on magnetization reversal. Owing to the different easy directions the grains switch at slightly different values of the external field. The coercive field is $\mu_0 H_\mathrm{c} = 4.77$~T. With $\alpha_K = 1$, $N_\mathrm{eff} = 0$, and $H_\mathrm{f} = 0$ which hold for case 1 per definition, we obtain $\alpha_\psi = 0.71$ from equation (\ref{eq_kroni}). For the computation of the dashed line (case 2) we {assume} a weakly ferromagnetic grain boundary phase. The magnetostatic terms are not taken into account. The grain boundary phase acts as a soft magnetic defect and reduces coercivity. The corresponding microstructural parameter is $\alpha_K = 0.67$. Owing to exchange coupling between the grains all grains reverse at the same external field. For the dotted line (case 3) we {switch} on the demagnetizing field. The magnetization and the demagnetizing field are shown in a slice through the grains in figures \ref{fig_microEffects}c and \ref{fig_microEffects}d, respectively. The reduction of the coercive field owing to demagnetizing effects equates to $N_\mathrm{eff} = 0.2$. Finally, we {take} into account thermal activation by computing the energy barrier for the nucleation of reversed domains as function of field \cite{bance2015thermal}. The decrease of coercivity by thermal activation {is} $\mu_0 H_\mathrm{f} = 0.23$~T.    

\section{Micromagnetics of permanent magnets}

\subsection{Micromagnetic energy contributions}

Micromagnetism is a continuum theory {that handles} magnetization processes on a length scale that is small enough to resolve the transition of the magnetization within domain walls but large enough to replace the atomic magnetic moments by a continuous function of position \cite{brown1959micromagnetics}. The state of the magnet is described by the magnetization $\mathbf M$, whose magnitude $\left| \mathbf M \right | = M_\mathrm{s}$ is constant and whose direction $\mathbf m = \mathbf M/M_\mathrm{s}$ is continuous. A stable or metastable magnetic state can be found by finding a function $\mathbf m = \mathbf m(\mathbf r)$ with $\left| \mathbf m \right| = 1$ that minimizes the Gibbs free energy of the magnet
\begin{eqnarray}
\label{eq_exchange}
E(\mathbf m)  = & &\int_V \left[ A\left(  \left( \nabla m_x \right)^2 + 
                           \left( \nabla m_y \right)^2 +
                           \left( \nabla m_z \right)^2 \right) \right. \\
\label{eq_ani}
      & & \qquad  - K \left( \mathbf m \cdot \mathbf k \right)^2 \\
\label{eq_magnetostatic}
      & & \qquad  - \frac{\mu_0 M_\mathrm{s}}{2}\left( \mathbf m \cdot \mathbf H_\mathrm{d} \right)  \\
\label{eq_zeeman}
      & & \Big. \qquad - \mu_0 M_\mathrm{s} \left( \mathbf m \cdot \mathbf H_\mathrm{ext} \right)  \Big] dV.
\end{eqnarray}
The different lines describe the exchange energy, the magnetocrystalline anisotropy energy, the magnetostatic energy, and the Zeeman energy, respectively. The coefficients $A$, $K$, and $M_\mathrm{s}$ in equation (\ref{eq_exchange}) to (\ref{eq_zeeman}) vary with position and thus represent the microstructure of the magnet. The unit vector along the anisotropy direction, $\mathbf k$, varies from grain to grain reflecting the orientation of the grains. The anisotropy constant $K$ will be zero in local defects or within the grain boundary phase. The grain boundary phase may be weakly ferromagnetic with magnetization $M_\mathrm{s}$ and the exchange constant $A$ considerably reduced with respect to the bulk values. In $\alpha$-Fe inclusions $K$ is negligible and $M_\mathrm{s}$ and $A$ are high. Composite magnets combine grains with different intrinsic properties. The demagnetizing field $\mathbf H_\mathrm{d}$ arises from the divergence of the magnetization. The factor 1/2 in equation (\ref{eq_magnetostatic}) indicates that it is a self-energy which depends on the current state of $\mathbf M$. It can be calculated from the static Maxwell's equations. One common method is the numerical solution of the magnetostatic boundary value problem for the magnetic scalar potential $U$ where the demagnetizing field is derived as $\mathbf H_\mathrm{d} = - \nabla U$. The magnetic scalar potential fulfills the Poisson equation
\begin{equation}
\label{eq_poisson}
\nabla^2 U = \mathbf \nabla \cdot \mathbf M
\end{equation}
inside the magnet, the Laplace equation
\begin{equation}
\label{eq_laplace}
\nabla^2 U = 0
\end{equation}
outside the magnet, and the interface conditions
\begin{eqnarray}
\label{eq_value}
U^\mathrm{(in)} = U^\mathrm{(out)} \\
\label{eq_gradient}
\left( U^\mathrm{(in)} - U^\mathrm{(out)}\right) \cdot \mathbf n = \mathbf M \cdot \mathbf n
\end{eqnarray}
at the magnet's boundary with unit surface normal $\mathbf n$. Equation (\ref{eq_value}) follows from the continuity of the component of the magnetic field $\mathbf H$ parallel to the surface (which follows from $ \mathbf \nabla \times \mathbf H = 0$). Equation (\ref{eq_gradient}) follows from the continuity of the component of the magnetic induction $\mathbf B$ normal to the surface (which follows from $ \nabla \cdot \mathbf B = 0$) \cite{Jackson}. 

\subsection{Numerical methods}

\subsubsection{Hysteresis}

There is no unique constrained minimum for equation (\ref{eq_exchange}) to (\ref{eq_zeeman}) for a given external field. The magnetic state that a magnet can access depends on its history. Hysteresis in a non-linear system results from the path formed by subsequent local minima \cite{kinderlehrer1997hysteretic}. In permanent magnet studies we are interested in the demagnetization curve. Thus, we use the saturated state as initial state and compute subsequent energy minima for a decreasing applied field, $H_\mathrm{ext}$. The projection of the magnetization onto the direction of the applied field integrated over the {volume of the} magnet, that is $\int_V M_\mathrm{s} \left( \mathbf m \cdot \mathbf H_\mathrm{ext}/ \left| \mathbf H_\mathrm{ext} \right| \right) dV $, as function of different values of $H_\mathrm{ext}$ gives the $M(H_\mathrm{ext})$-curve. For computing the maximum energy density product we need the $M(H)$-curve, where $H$ is the internal field $H = H_\mathrm{ext} - N M (H_\mathrm{ext})$. Similar to open circuit measurements \cite{cullity2011introduction} we correct  $M(H_\mathrm{ext})$ with the macroscopic demagnetizing factor $N$ of the sample, in order to obtain $M(H)$.

\subsubsection{Finite element and finite difference discretization}

The computation of the energy for a permanent magnet requires the discretization of equation{s} (\ref{eq_exchange}) to (\ref{eq_zeeman}) taking into account the local variation of $M_\mathrm{s}$, $K$, and $A$, according to the microstructure. Common discretization schemes used in micromagnetics for permanent magnets \cite{fidler2000micromagnetic} are the finite difference method \cite{tsukahara2016large} or the finite element method \cite{tanaka2017speeding,fischbacher2017nonlinear}. 

Each node of a finite element mesh or cell of a finite difference scheme with index $i$ holds a unit magnetization vector $\mathbf m_i$. We gather these vectors into the vector $\mathbf{x}$ which has the dimension $3n$, where $n$ is the number of nodes or cells. Then the Gibbs free energy may be written as \cite{fischbacher2017nonlinear}
\begin{equation}
\label{eq_discrete}
E(\mathbf x) = \frac{1}{2} \mathbf{x}^{\mathrm T} \mathbf{C} \mathbf{x} - \frac{\mu_0}{2}\mathbf{h}_{\mathrm d}^{\mathrm T} \bar{\mathbf{M}} \mathbf{x} - \mu_0 \mathbf{h}_{\mathrm{ext}}^{\mathrm T} \bar{\mathbf{M}} \mathbf{x}. 
\end{equation}
The three terms on the right-hand side of (\ref{eq_discrete}) from left to right are the sum of the exchange and anisotropy energy, the magnetostatic self-energy, and the Zeeman energy, respectively. The sparse matrix $\mathbf C$ contains grid information associated with the {discretization of the} exchange and anisotropy {energy}. The matrix $\bar{\mathbf{M}}$ accounts for the local variation of the saturation magnetization $M_\mathrm{s}$ within the magnet. It is a diagonal matrix whose entries are the modulus of the magnetic moment associated with the node or cell $i$ \cite{fischbacher2017nonlinear}. The vectors $\mathbf x$, $\mathbf{h}_{\mathrm d}(\mathbf{x})$, and $\mathbf{h}_{\mathrm{ext}}$ hold the unit vectors of the magnetization, the demagnetizing field, and the external field at the nodes of the finite element mesh or the cells of a finite difference grid, respectively. 

For computing the demagnetizating field, equations (\ref{eq_poisson}) to (\ref{eq_gradient}) can be solved using an algebraic multigrid method on the finite element mesh \cite{sun2006adaptive,shepherd2014discretization,fischbacher2017nonlinear}.

In finite difference methods the magnetization is assumed to be uniform within each cell. Then the magnetic field generated at point $\mathbf r$ by the magnetization in cell $j$ is given by an integration over the magnetic surface charges $\sigma_j = M_{\mathrm{s}j}\mathbf{m}_j \cdot \mathbf n$ \cite{aharoni2000introduction},
\begin{equation}
\label{eq_fieldfromcell}
\mathbf H_{\mathrm{d},j}(\mathbf r) = - \frac{1}{4\pi} \nabla \left( \int_{\partial V_j} \frac{ \sigma_j'}{\left| \mathbf r - \mathbf r' \right|} dS_j' \right),
\end{equation}
where $\mathbf n$ is the unit surface normal.  The magnetostatic energy is a double sum over all computational cells
\begin{equation}
\label{eq_emag}
E_\mathrm{m} = -\frac{\mu_0}{2} \sum_i
M_\mathrm{s,i}\mathbf m_i \cdot
\int_{V_i}    \sum_j \mathbf H_{\mathrm{d},j}(\mathbf r)  dV.
\end{equation}
Applying integration by parts we can rewrite the magnetosatic energy as \cite{fabian1996three}
\begin{equation}
\label{eq_emagsurfsurf}
E_\mathrm{m}  =  \frac{\mu_0}{8\pi} \sum_{i,j}  \int_{\partial V_i} \int_{\partial V_j}
\frac{ \sigma_i \sigma_j'}
     { \left| \mathbf r - \mathbf r' \right|} dS_j' dS_i.
\end{equation}
Introducing the demagnetization tensor $\mathbf N_{ij}$  reduces equation (\ref{eq_emagsurfsurf}) to
\begin{equation}
\label{eq_emagwithtensor}
E_\mathrm{m}  =  \frac{\mu_0}{2} V_\mathrm{cell} \sum_{i,j} M_{\mathrm{s}i}\mathbf m_i \mathbf N_{ij} M_{\mathrm{s}j}\mathbf m_j,
\end{equation}
where $V_\mathrm{cell}$ is the volume of a computational cell. The term $\mu_0 V_\mathrm{cell}M_{\mathrm{s}i}\mathbf m_i \mathbf N_{ij} M_{\mathrm{s}j} \mathbf m_j $
is the magnetostatic interaction energy between cells $i$ and $j$. From equation (\ref{eq_emagwithtensor}) we can compute the cell averaged demagnetizing field \cite{schabes1987magnetostatic,nakatani1989direct}
\begin{equation}
\label{eq_conv}
\mathbf h_i = - \sum_j \mathbf N_{ij} M_{\mathrm{s}j}\mathbf m_j.
\end{equation}
The demagnetization tensor $\mathbf N_{ij}$ depends only on the relative distance between the cells $i$ and $j$. The convolution (\ref{eq_conv}) can be efficiently computed using Fast Fourier Transforms.
Special implementations of the Fast Fourier Transform with low communication overhead makes large-scale simulations of permanent magnets possible on supercomputers with thousands of cores \cite{tsukahara2016large}.

\subsubsection{Energy minimization}

The intrinsic time scale of magnetization processes is related to the Lamor frequency $f = \gamma \mu_0 H/(2\pi)$. The gyromagnetic ratio is $\gamma = 1.76086 \times 10^{11}$/(Ts). For example, let us estimate the intrinsic time scale for precession in Nd$_2$Fe$_{14}$B. The magnitude of typical internal fields, $\mu_0 H$ are about a few Tesla. The Lamor frequency is 28 GHz per Tesla. This gives a characteristic time scale smaller than $10^{-10}$~s. Such time scales may be relevant for magnetic recording or spin electronic devices. In permanent magnet applications the rate of change of the external field is much slower. {For example, low speed direct drive wind mills run at about 20 rpm \cite{pavel2017substitution} and  motors of hybrid vehicles run at 1500 rpm to 6000 rpm \cite{dorrell2010comparison}, which translate into frequencies ranging from 1/3~Hz to 100~Hz.}  The magnetization always reaches metastable equilibrium before a significant change of the external field. Based on this argument  many researcher{s} use energy minimization methods for simulation of magnetization reversal in permanent magnets{,} taking advantage of a significant {speedup} as compared to time integration solvers \cite{tanaka2017speeding,exl2014labonte}.

Minimizing equation (\ref{eq_discrete}) subject to the unit norm constraint for decreasing external field gives the magnetic states along the demagnetization curve of the magnet. The sparse matrix $\mathbf C$ and the diagonal matrix $\bar{\mathbf M}$ depend only on the geometry and the intrinsic magnetic properties. The vector $\mathbf h_\mathrm{d}$ depends linearly on the magnetization. Evaluation of the energy, its gradient, or the Hessian requires {the solution of} the magnetostatic subproblem. The magnetic field depends linearly on the magnetization. Thus, (\ref{eq_discrete}) is quadratic in $\mathbf x$. However, the condition $\mathbf M = M_\mathrm{s}$ implies that each subvector $\mathbf m_i$ of $\mathbf x$ is constrained to be a unit vector. The optimization problem is supplemented by  $n$ nonlinear constraints $\left| \mathbf m_i \right| = 1, \; i = 1,...,n$. 

In its most simple form an algorithm for energy minimization (see algorithm \ref{alg_mini}) is an iterative process with the following four tasks per iteration: The computation of the search direction, the computation of the step length, the motion towards the minimum, and the check for convergence. These computations make use of the objective function $E(\mathbf x)$ and its gradients, possibly second derivatives and maybe information gathered from previous iterations. The superscript $^+$ marks quantities which are computed for the next iteration step. The superscript $^-$ marks quantities which have been computed during the previous iteration step.

\begin{algorithm}
\caption{minimize $E(\mathbf x)$}
\label{alg_mini}
\begin{algorithmic}
\Repeat
\State compute search direction $\mathbf d$
\State compute step length $\alpha$
\State proceed $\mathbf x^+ = \mathbf x + \alpha \mathbf d$
\Until{convergence}
\end{algorithmic}
\end{algorithm}

Variants of the steepest-descent method, \cite{exl2014labonte,furuya2015semi,oikawa2016large,ke2017simulation}, the non-linear conjugate gradient method \cite{fischbacher2017nonlinear,tanaka2017speeding}, and the quasi-Newton method \cite{sepehri2017correlation,sodervznik2017magnetization,tang2018coercivity} are most widely used in micromagnetics for permanent magnets.
In steepest-descent methods the search direction is the negative gradient $\mathbf g = \nabla E(\mathbf x)$ of the energy: $\mathbf d = -\mathbf g$. The nonlinear conjugate-gradient method uses a sequence of conjugate directions $\mathbf d = - \mathbf g + \beta \mathbf d^{-}$.  The Newton method uses the negative gradient multiplied by the inverse
of the Hessian matrix: $\mathbf d = - (\nabla^2 E)^{-1} \nabla \mathbf g$ as search direction.
In quasi-Newton methods the inverse of the Hessian is approximated by
information gathered from previous iterations.

The step length $\alpha$ is obtained by approximate line search minimization. To that end the new point is determined along the line defined by the current search direction and should yield a sufficiently smaller energy with {a} sufficient{ly} small gradient  (approximate local minimum along the line).
By expanding and shrinking a search interval for $\alpha$, line search algorithms \cite{more1994line} find an appropriate step length. Owing to the solution of the magnetostatic subproblem, evaluations of the energy are expensive. In order to reduce the number of energy evaluations Koehler and Fredkin \cite{koehler1992finite} apply an inexact line search based on cubic interpolation. Tanaka and co-workers \cite{tanaka2017speeding} propose to interpolate the magnetostatic field within the search interval if it is sufficiently small. Fischbacher et al. \cite{fischbacher2017nonlinear} showed that long steps {should} be avoided, in order to compute all metastable states along the demagnetization curve. They suggest applying a single Newton step in one dimension to get an initial estimate for the step length which then may be further reduced to fulfill the sufficient decrease condition. For steepest descent methods Barzilai and Borwein \cite{barzilai1988two} proposed a step length $\alpha$ such that $\alpha$ multiplied with the identity matrix, $\mathbf 1$, approximates the inverse of the Hessian matrix: $\alpha \mathbf 1 \approx  (\nabla^2 E)^{-1}$. Thus, the Barzilai-Borwein method makes use of the key idea of limited memory quasi Newton methods applied to step length selection. The step length is computed from information gathered during the last 2 iteration steps. This method was successfully applied in numerical micromagnetics \cite{exl2014labonte,abert2014efficient,vansteenkiste2014design}.

When applied {to} micromagnetics the update rule $\mathbf x^+ = \mathbf x + \alpha \mathbf d$ will not preserve the norm of the magnetization vector. A simple cure is renormalization $\mathbf m^+_i = (\mathbf m_i + \alpha  \mathbf d_i) / \left| (\mathbf m_i + \alpha  \mathbf d_i) \right|$. When certain conditions for $\mathbf{d}$  \cite{alouges1997new} and the computational grid  \cite{bartels2005stability} are met the normalization leads to a decrease in the energy. One condition  \cite{alouges1997new} for an energy decrease upon normalization is that the search direction is perpendicular to the magnetic state of the current point: $\mathbf m_i \cdot \mathbf d_i$ for all $i$. Following Cohen et al. \cite{cohen1989relaxation} we can replace the energy gradient by its projection on{to} its component perpendicular to the local magnetization
\begin{equation}
\label{eq_project}
\hat{\mathbf g}_i = \mathbf g_i - \left( \mathbf g_i \cdot \mathbf m_i  \right) \mathbf m_i = - \mathbf m_i \times \left( \mathbf m_i \times \mathbf g_i \right).
\end{equation}
In nonlinear conjugate gradient methods the search directions are linear combinations of vectors perpendicular to the magnetization (the current projected gradient and the previous search directions initially being the projected gradient). Instead of updating and normalization, the vectors $\mathbf m_i$ might also be rotated by an angle $\alpha \left|\mathbf d_i \right|$ \cite{tsukahara2017magnetization} or a norm conserving semi-implicit update rule \cite{goldfarb2009curvilinear} may be applied.  

The right-hand side of equation (\ref{eq_project}) follows from the vector identity 
$\mathbf a \times \left( \mathbf b \times \mathbf c\right) = \left( \mathbf a \cdot \mathbf c \right) \mathbf b - \left( \mathbf a \cdot \mathbf b \right) \mathbf c$ and 
$\left| \mathbf m_i \right| = 1$. On computational grids with {a} uniform mesh the energy gradient is proportional to the effective field 
\begin{equation}
\label{eq_heff}
\mathbf h_{\mathrm{eff}i} =  - \frac{1}{\mu_0 M_{\mathrm{s}i} V_\mathrm{cell}} \mathbf g_i. 
\end{equation}
By inspecting the right-hand side of equation (\ref{eq_project}), we see that the search direction of a steepest-descent method is proportional to the damping term of the Landau-Lifshitz equation \cite{landau1935theory}. Time integration of the Landau-Lifshitz-Gilbert equation without the precession term \cite{furuya2015semi} 
is equivalent to minimization by the steepest-descent method. Furaya et al. \cite{furuya2015semi} use a semi-implicit time integration scheme. They split the effective field into its local part and the long-ranging magnetostatic field which not only depends on the nearest neighbor cells but on the magnetization in the entire magnet. By treating the local part of the effective field implicitly and the magnetostatic field explicitly, much larger time steps and thus faster convergence toward the energy minimum is possible.   
 
There are several possible termination criteria for a minimization algorithm. Koehler and Fredkin \cite{koehler1992finite} used the relative change in the energy between subsequent iterations. Others \cite{furuya2015semi,tsukahara2016large} use the difference between the subsequent magnetic states. Gill et al. \cite{gill1981practical} recommend a threefold criterion taking into account the change in energy, the change in the magnetic state, and the norm of the gradient for unconstrained optimization. This ensures convergence of the sequence of the magnetic states, avoids early stops in flat regions and ensures progress towards the minimum.  

\subsubsection{Time integration}

The torque on the magnetic moment $\mathbf M V_\mathrm{cell}$ of a computational cell in a magnetic field $\mathbf H$ is $\mathbf T = \mu_0 \mathbf M V_\mathrm{cell} \times \mathbf H$. The angular momentum associated with the magnetic moment is $\mathbf L = -\mathbf M V_\mathrm{cell}/\left|\gamma\right|$. The change of the angular momentum with time equals the torque, $\partial \mathbf L / \partial t = \mathbf T$. Applying the torque equation for the magnetic volume which is divided into computational cells gives 
\begin{equation}
\label{eq_pre}
\frac{\partial \mathbf m_i }{ \partial t} = -\left| \gamma \right| \mu_0 \mathbf m_i \times \mathbf h_{\mathrm{eff}i},
\end{equation}
which describes the precession of the magnetic moment around the effective field. In order to describe the motion of the magnetization towards equilibrium, equation (\ref{eq_pre}) has to be augmented with a damping term. Following Landau and Lifshitz \cite{landau1935theory} we can add a term $-\lambda \mathbf m_i \times \left( \mathbf m_i \times \mathbf h_{\mathrm{eff}i} \right)$ to the right-hand side of equation (\ref{eq_pre}) which will move the magnetization towards the field. Alternatively we can - as suggested by Gilbert \cite{gilbert1955lagrangian} - add a dissipative force $-\alpha \partial \mathbf m_i / \partial t$ to the effective field. The precise path the system {follows} towards equilibrium depends on the type of equation used and the value of the damping parameters $\lambda$ or $\alpha$. In the Landau-Lifshitz equation precession is not changed with increasing damping, whereas in the Gilbert case an increase of the damping constant slows down precessional motion. Only for small damping the Landau Lifshitz equation and the Gilbert equation are equivalent. This can be seen if the Gilbert equation is written in Landau-Lifshitz form \cite{mallinson1987damped}
\begin{eqnarray}
\label{eq_llg}
\frac{\partial \mathbf m_i }{ \partial t} & = & - \frac{\left| \gamma \right| \mu_0}{1+\alpha^2} \mathbf m_i \times \mathbf h_{\mathrm{eff}i} \nonumber \\ 
                                   & & - \alpha \frac{\left| \gamma \right| \mu_0}{1+\alpha^2}
                                    \mathbf m_i \times \left( \mathbf m_i \times \mathbf h_{\mathrm{eff}i} \right).
\end{eqnarray}
In the limit of high damping only the second term of equation (\ref{eq_llg}) remains and the time integration of the Landau-Lifshitz-Gilbert equation becomes equivalent to the steepest-descent method. For coherent rotation of the magnetization the minimum reversal time occurs for a damping parameter $\alpha = 1$. In turn, fast reversal reduces the total computation time. This is the motivation for using a damping parameter $\alpha = 1$ for simulation of magnetization reversal in permanent magnets \cite{tsukahara2017micromagnetic,tsukahara2017magnetization} by numerical integration of equation (\ref{eq_llg}). Several public domain micromagnetic software tools use solvers for the numerical solution of equation(\ref{eq_llg}) based on the adaptive Euler methods \cite{fu2016finite}, Runge-Kutta schemes \cite{vansteenkiste2014design}, backward-differentiation methods \cite{fangohr2016nmag}, and preconditioned implicit solvers \cite{scholz2003scalable}.

\subsubsection{Energy barriers}

\begin{figure}
	\includegraphics[width=8.0cm]{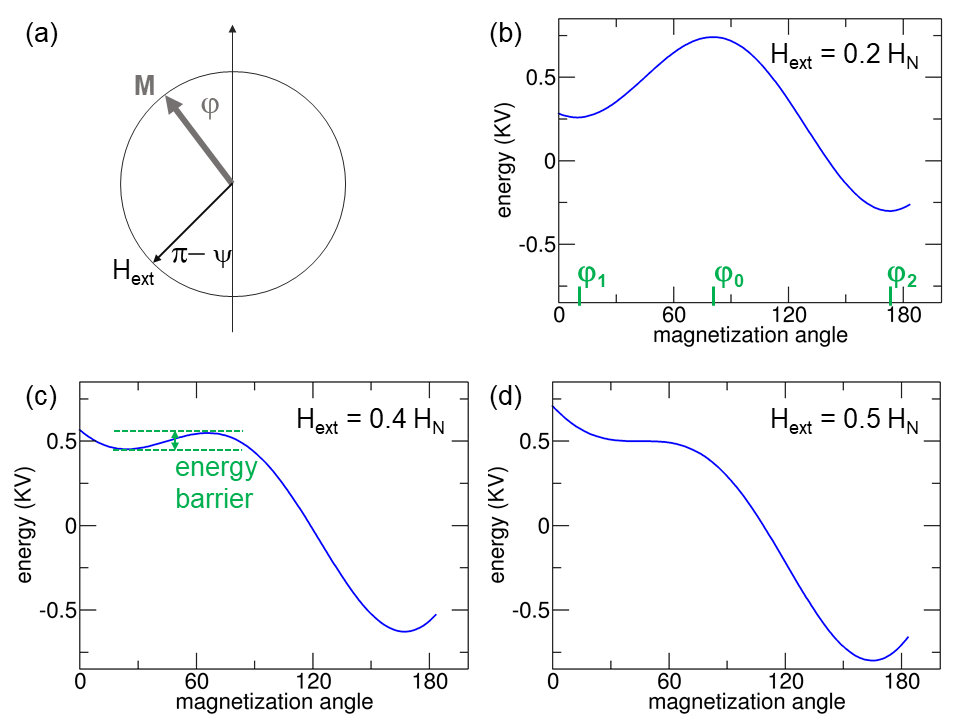}
	\caption{\label{fig_sw}Energy landscape as function of the magnetization angle. (a) Small hard magnetic sphere in an external field. The field is applied at an angle of $\pi-\psi = 45$ degrees. (b) to (d) With increasing field the energy barrier decreases.}
\end{figure}

Permanent magnets are used at elevated temperature. However, classical micromagnetic simulations take into account temperature only by the temperature-dependent intrinsic materials properties. Thermal fluctuations that may drive the system over a finite energy barrier are neglected. Before magnetization reversal, a magnet is in a metastable state. With increasing opposing field the energy barrier decreases. The system follows the local minima reversibly until the energy barrier vanishes and the magnetization changes irreversibly \cite{schabes1991micromagnetic}. If the height of the energy barrier is around $25 k_\mathrm{B}T$, thermal fluctuations can drive the system over the barrier within a time of approximately one second \cite{gaunt1976magnetic}. To illustrate this behavior, let us look at the energy landscape of a small hard magnetic sphere with volume $V$ (see figure \ref{fig_sw}). The energy per unit volume is $E(\varphi,H_\mathrm{ext})/V = K \sin^2(\varphi) - \mu_0 M_\mathrm{s} H_\mathrm{ext} \cos(\varphi-\psi)$. The external field is applied at an angle $\psi$ with respect to the positive anisotropy axes. For small external fields the energy shows two minima as function of the magnetization angle $\varphi$. The state before switching is given by $\varphi_1 = (\pi-\psi)M_\mathrm{s}H_\mathrm{ext}/(2K-M_\mathrm{s}H_\mathrm{ext})$ and the state after switching is given by $\varphi_2 = \pi - (\pi-\psi)M_\mathrm{s}H_\mathrm{ext}/(2K+M_\mathrm{s}H_\mathrm{ext})$. The maximum energy occurs at the saddle point at $\varphi_0 = \sqrt[3]{-\tan \psi}$ \cite{kronmuller1987angular}.      

\begin{figure}
\includegraphics[width=8.0cm]{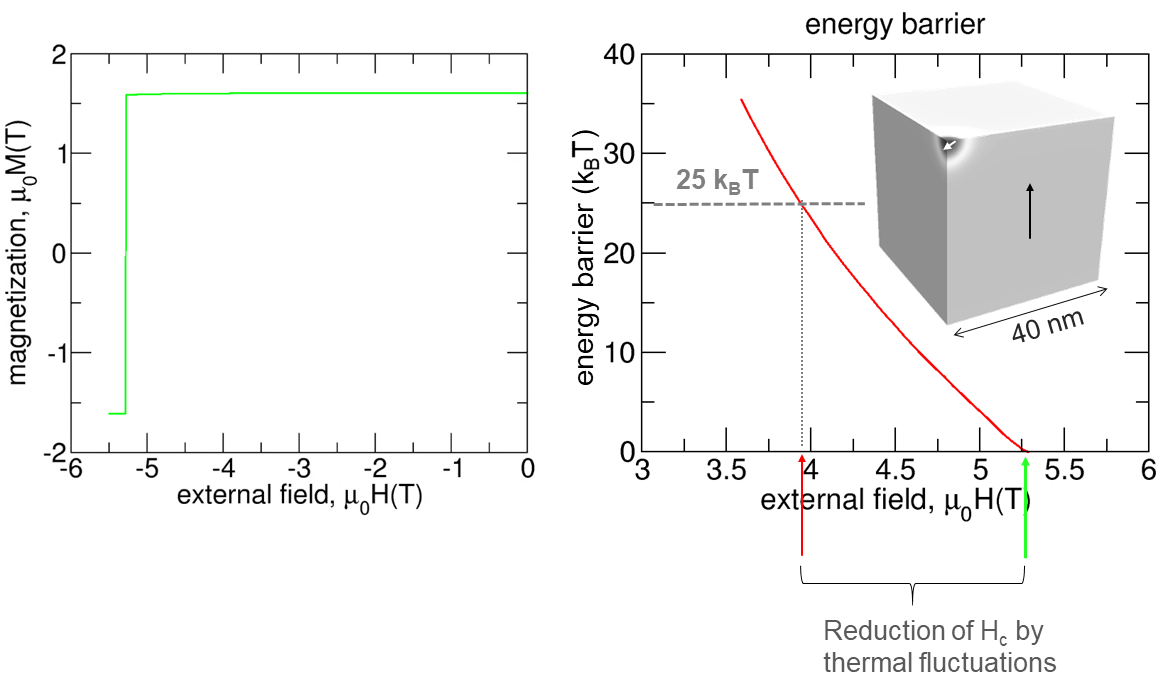}
\caption{\label{fig_barrier}Thermally induced magnetization reversal in a Nd$_2$Fe$_{14}$B cube. Left: Computed demagnetization curve by classical micromagnetics. Right: Energy barrier for the formation of a reversed nucleus as {a} function of the applied field. The inset shows the saddle point configuration of the magnetization. {Data taken from \cite{fischbacher2017limits}}.}
\end{figure}

In permanent magnets magnetization reversal occurs by the nucleation and expansion of reversed domains \cite{givord1992coercivity}. Similar to the situation depicted in figure \ref{fig_sw} the nucleation of a reversed domain or the depinning of a domain wall is associated with an energy barrier that is decreased by an increasing external field. Using the elastic band method \cite{dittrich2002path} or the string method \cite{weinan2002string} the energy barrier can be computed as function of the external field. The critical field at which the energy barrier $E_\mathbf{B}(H_\mathrm{ext})$ crosses the $25 k_\mathrm{B}T$-line is the coercive field of the magnet taking into account thermal fluctuations. 
The elastic band method and the string method are well-established path finding methods in chemical physics \cite{henkelman2000improved,zhang2016recent}. In micromagnetics they can be used to compute the minimum energy path that connects the local minimum at field $H_\mathrm{ext}$ with the reversed magnetic state. A path is called a minimum energy path if for any point along the path the gradient of the energy is parallel to the path. In other words, the component of the energy gradient normal to the path is zero. The string method can be easily applied by subsequent application of a standard micromagnetic solver. It is an iterative algorithm: The magnetic states along the path are described by images. Each image is a replica of the total system.  A single iteration step consists of two moves. (1) Each image is relaxed by applying a few steps of an energy minimization method or by integrating the Landau-Lifshitz-Gilbert equation for a very short time. (2) The images are moved along the path such that the distance between the images is constant. Within the framework of the elastic band method images may only move perpendicular to the current path and the distance between the images is kept constant with a virtual spring force. For an accurate computation of the energy barrier for a nucleation process \cite{zhang2016recent} variants of the string method exists which keep more images next to the saddle point. This can be achieved by an energy weighted distance function between the images \cite{weinan2007simplified} and truncation of the path \cite{carilli2015truncation}. 

Figure \ref{fig_barrier} compares the coercive field of a Nd$_2$Fe$_{14}$B cube with an edge length of 40~nm obtained by classical micromagnetic simulations and computing energy barriers as discussed above. In both methods the intrinsic magnetic material parameters for $T = 300$~K were used. The non-zero temperature coercive field, which takes into account thermal fluctuations, is defined as the critical value of the external field at which the energy barrier reaches $25 k_\mathrm{B}T$. By inspecting the magnetic states along the minimum energy path we can see how thermally induced magnetization reversal happens. At the saddle point a small reversed nucleus is formed. If there is no barrier for the expansion of the reversed domain, the reversed domain grows and the particle will switch. The simulations are self-consistent: The coercive field calculated by classical micromagnetics equals the field at which the energy barrier vanishes. For nearly ideal particles such as the cube without soft magnetic defects discussed above, the reduction of the coercive field by thermal fluctuations may be as large as 25 percent \cite{bance2015thermally}. However, the presence of defects reduce{s} the decay of coercivity owing to thermal fluctuations \cite{fischbacher2017limits}. For example for the magnetic structure {in} figure \ref{fig_HcDy}{,} which contains a weakly ferromagnetic grain boundary, thermal fluctuations reduce the coercive field by only 8 percent.  

Energy barriers for reversal may also be computed by atomstic spin dynamics. Miyashita et al. \cite{hirosawa2017perspectives,miyashita2017perspectives} solved equation (\ref{eq_llg}) numerically for atomic magnetic moments augmented by a stochastic thermal field. From the computed relaxation time, $\tau$, at a fixed external field the energy barrier can be computed by fitting the results to an Arrhenius-Neel law $\tau = (1/f_0) \exp(E_\mathrm{B}/(k_\mathrm{B}T))$ or to Sharrock's law \cite{sharrock1994time}, which gives the coercive field as function of $H_\mathrm{ext}$ and $\tau$. Alternatively, Toga et al. used the constrained Monte-Carlo method \cite{toga2016monte} to compute field dependent energy barriers for an atomistic spin model.

In equation (\ref{eq_kroni}) we {attributed} the reduction of coercivity to the fluctuation field $H_\mathrm{f}$. The energy barrier for magnetization reversal is related to {this} fluctuation field by $H_\mathrm{f} = - 25 k_\mathrm{B}T/(\partial E / \partial H_\mathrm{ext})$. Experimentally, the energy barrier or the fluctuation field can be obtained by measuring the magnetic viscosity which is related to the change of magnetization with time at a fixed external field. It was measured by Givord et al. \cite{givord1987magnetic}, Villas-Boas \cite{villas1998magnetic}, and Okamota et al. \cite{okamoto2015temperature} for sintered, melt-spun, and hot-deformed magnets, respectively.         

\subsection{Microstructure representation}

\begin{figure}
\includegraphics[width=8.0cm]{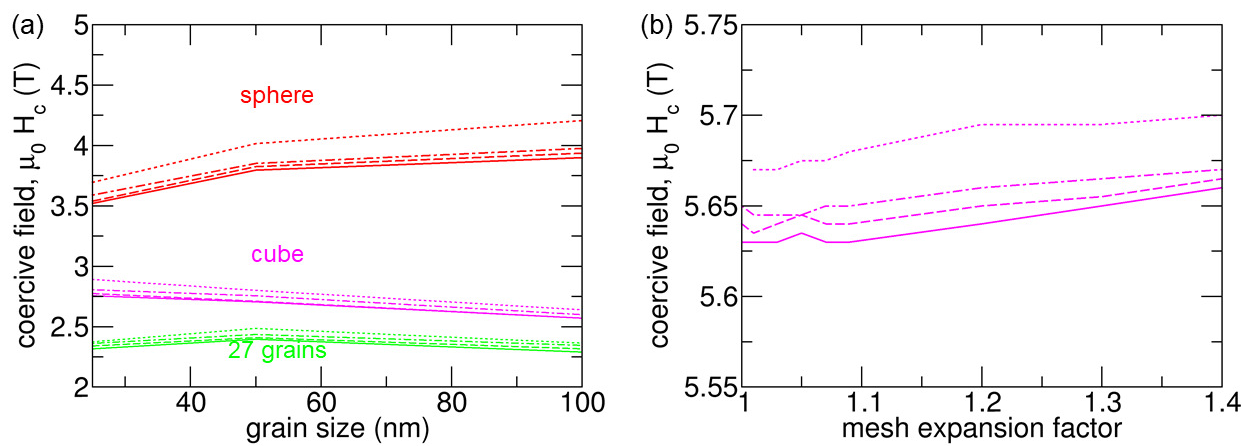}
\caption{\label{fig_mesh} (a) Coercive field of a sphere, a cube, and a polycrystalline magnet as function of grain size. The grains are made of Nd$_2$F$_{14}$B and surrounded by a 3 nm  thick, weakly-ferromagnetic grain boundary phase. The grain size is defined as $\sqrt[3]{V_\mathrm{grain}}$, where $V_\mathrm{grain}$ is the volume of the grain. (b) Coercive field of the 100~nm pure Nd$_2$F$_{14}$B cube as function of the expansion factor $a$ in a geometrical mesh. The solid, dashed, dot-dashed, and dotted line refer to a mesh size of 1.3~nm, 2~nm, 2.7~nm, and 4~nm, respectively, defined at the surface of the cube{.}}
\end{figure}

\subsubsection{Grain size and particle shape}

The discretization of the Gibb's free energy by finite differences or finite elements {poses a} question concerning the required grid size. The required grid size is related to the characteristic length scale of inhomogeneities in the magnetization{,} which is related to the relative weight of the exchange energy to other contributions of the Gibb's free energy. 

Upon minimization the exchange energy favors a uniform magnetization with the local magnetic moments on the computational grid parallel to each other. The accurate computation of the critical field for the formation of a reversed nucleus requires the energy of the domain wall, which separates the nucleus from the rest, {to be known} with high accuracy. Therefore, we should be able to resolve the transition of the magnetization within the domain wall on the computational grid. The width of a Bloch wall is $\delta_\mathrm{B} = \pi \delta_0$. The Bloch wall parameter $\delta_0 = \sqrt{A/K}$ denotes the relative importance of the exchange energy versus crystalline anisotropy energy. 

Whereas in ellipsoidal particles the demagnetizing field is uniform, it is inhomogeneous in polyhedral particles. The non-uniformity of the demagnetizing field strongly influence magnetization reversal \cite{schabes1988magnetization}. Near edges or corners \cite{gronefeld1989calculation} the transverse component of the demagnetizing field diverges. Owing to the locally increased demagnetizing field, the reversed nucleus will form near edges or corners \cite{thielsch2013dependence} (see also figure \ref{fig_barrier}). We have to {correctly} resolve the rotations of the magnetization that eventually form the reversed nucleus. For the computation of the nucleation field the required minimum mesh size has to be smaller than the exchange length $l_\mathrm{ex} = \sqrt{A/(\mu_0 M_\mathrm{s}^2/2)}$ at the place where the initial nucleus is formed. It gives the relative importance of the exchange energy versus magnetostatic energy. Please note that sometimes the exchange length is also defined as $l_\mathrm{lex} = \sqrt{A/(\mu_0 M_\mathrm{s}^2)}$ \cite{skomski2008simple,coey_2010}. In order to keep the computation time low and resolve important magnetization processes, Schmidts and Kronm\"uller introduced a graded mesh that is refined towards the edges \cite{schmidts1991size}. 

The relative importance of the different energy terms also explains the grain size dependence of coercivity. The coercive field {of} permanent magnets decrease{s} with increasing grain size \cite{schmidts1991size,bance2014grain,sepehri2014micromagnetic,liu2015grain,yazid2016mfm}. The smaller the magnet the more dominant is the exchange term. Thus, it costs more energy to form a domain wall. To achieve magnetization reversal, the Zeeman energy of the reversed magnetization in the nucleus needs to be higher. This can be accomplished by a larger external field.

In the following numerical experiment we computed the coercive field of a sphere, a cube, and a magnet consisting of 27 polyhedral grains. The polycrystalline magnet is shown in figure \ref{fig_poly}. We computed the coercive field as function of the size of the magnet for different finite element meshes. We used the conjugate gradient method \cite{fischbacher2017nonlinear} to compute the magnetic states along the demagnetization curve. The Nd$_2$Fe$_{14}$B particles ($K=4.9$~MJ/m$ ^3$, $\mu_0 M_\mathrm{s} = 1.61$~T, $A = 8$~pJ/m \cite{coey_2010}) were covered by a soft magnetic phase with a thickness of 3~nm. In the polycrystalline sample the grains are also covered by a 3 nm soft phase which adds up to a 6 nm thick grain boundary. The material parameters of the grain boundary phase $K=0$, $\mu_0 M_\mathrm{s} = 0.477$~T, and $A=6.12$~pJ/m correspond to a composition of Nd$_{40}$Fe$_{60}$ \cite{sakuma2015magnetism}. The characteristic lengths for the main phase are $\delta_0 = 1.3$~nm,  $\delta_\mathrm{B} = 4$~nm, and $l_\mathrm{ex} = 2.8$~nm. The exchange length for the boundary phase is $l_\mathrm{ex} = 8.2$~nm. The results are {summarized} in figure \ref{fig_mesh}a which give{s} the coercive field as function of grain size. The three sets of curves are for the sphere, the cube, and the polycrystalline magnet. The different curves within a set correspond to a mesh size of 1.3~nm, 2~nm, 2.7~nm, and 4~nm, defined at the surface of each grain and a mesh expansion factor of 1.05 for all models. As compared to the sphere the coercive field of the cube is reduced by about 1~T/$\mu_0$. For the cube the coercive field decreases with the particle size. In the sphere the demagnetizing field is uniform. The ratio of the hard (core) versus the soft phase (shell) determines coercivity. With increasing grain size the volume fraction of the soft phase decreases and coercivity increases. In all samples the coercive field decreases with decreasing mesh size. For all simulated cases, the relative change in the coercive field is less than two percent for a change of the mesh size from 1.3 nm to 2.7 nm.

In figure \ref{fig_mesh}b we present the results for the coercive field obtained by graded meshes. In a geometrical mesh \cite{szabo1991finite} the mesh size is gradually changed according to a geometric series. Towards the center of the grain the mesh size $h$ increases according to $h \times a^n$; where $a$ is the mesh expansion factor and $n$ is the distance to the surface measured by the number of elements. The coercive field increases with increasing $n$. However, for $a < 1.1$ there is almost no change in the coercivity. 
On the other hand, the number of finite element cells is reduced from 3.2 million for $a = 1.01$ to 1.6 million for $a = 1.09$ and a mesh size of 1.3 nm at the boundary. In this case, the runtime of the simulation was reduced by a factor 4, with both simulations computed on a single NVidia Tesla K80 GPU.
The situation is different if the cube contains a soft magnetic inclusion in the center which will act as nucleation site. Then a fine mesh is also required at the interface between the hard and the soft phase.
  
\subsubsection{Representation of multi-grain structures}

\begin{figure}
\includegraphics[width=8.0cm]{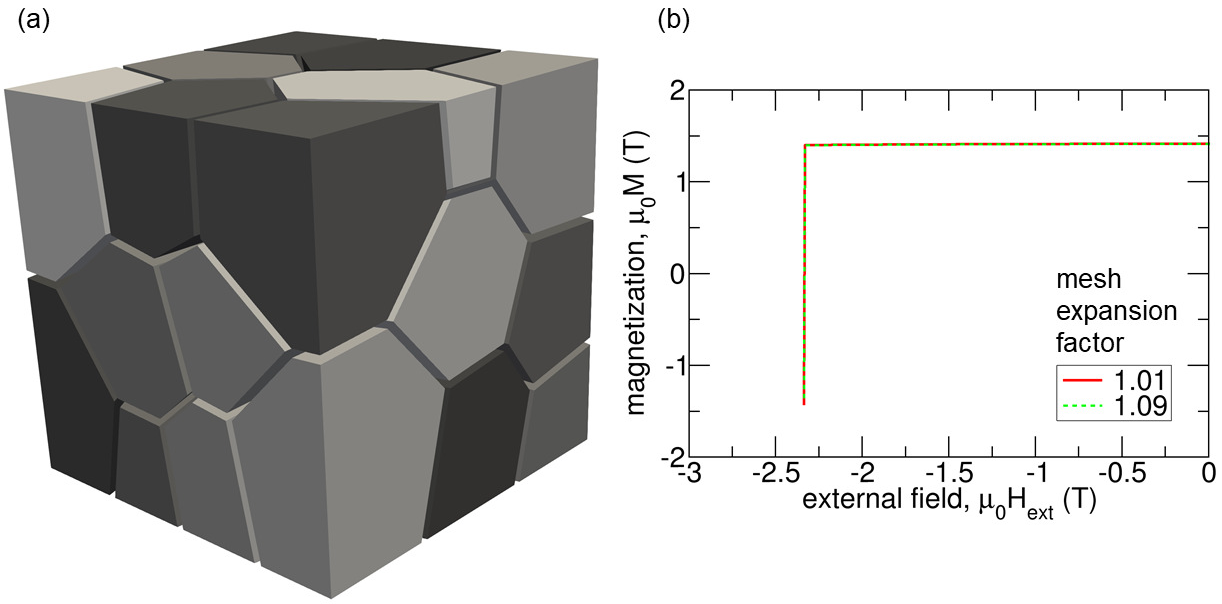}
\includegraphics[width=8.0cm]{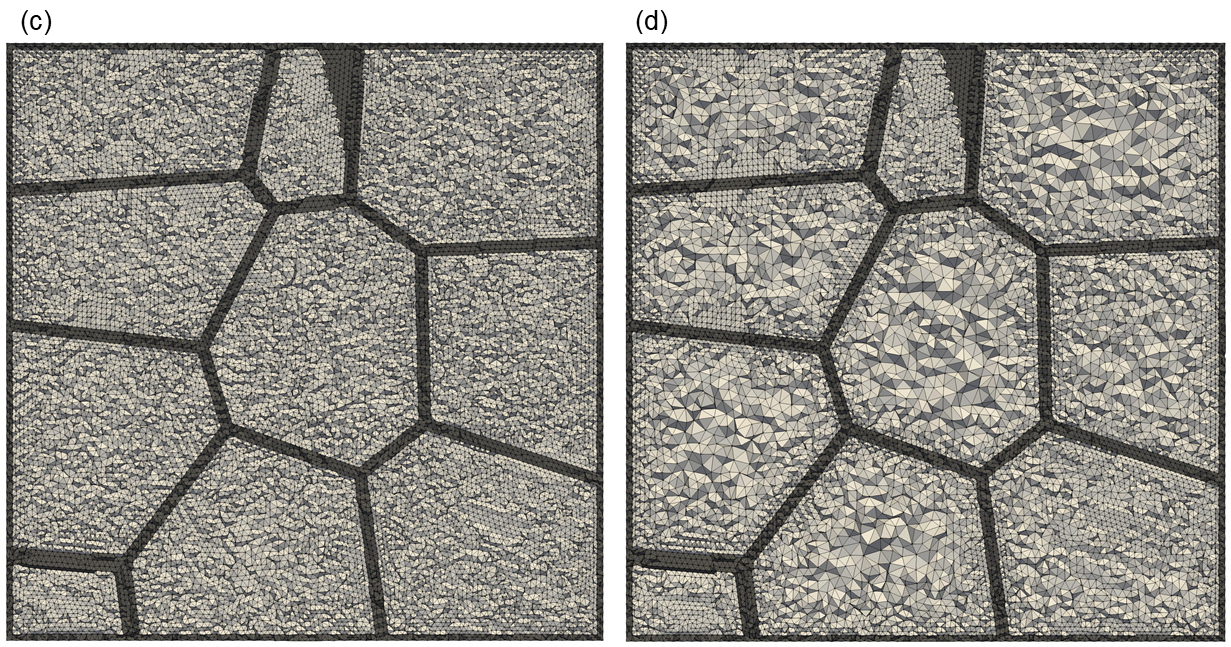}
\caption{\label{fig_poly}(a) Polycrystalline model of a Nd$_2$Fe$_{14}$B magnet. The edge length of the cube is 300~nm. (b) The computed demagnetization curve is the same for (c) an almost uniform mesh ($a = 1.01$) and (d) a geometric mesh with expansion factor $a=1.09$.}
\end{figure}

Computer programs for the semi-automatic generation of synthetic structures are essential to study the influence of the microstructure on the hysteresis properties of permanent magnets. Microstructure features that need to be taken into account are the properties of the grain
boundary phase \cite{sepehri2013high,zickler2015nanoanalytical,liu2016coercivity}, anisotropy enhancement by grain boundary diffusion \cite{bance2015thermally,oikawa2016large,helbig2017experimental,fischbacher2017searching,tang2018coercivity,li2018micromagnetic}, and the shape of the grains \cite{yi2016micromagnetic,sepehri2016micromagnetic,kovacs2017micromagnetic,erokhin2017optimization,fischbacher2017searching}. The grain boundary properties may be anisotropic based on the orientation of the grain boundary with respect to the anisotropy direction \cite{zickler2017combined,zickler2017nanocompositional,fujisaki2016micromagnetic}.

Software tools for the generation of synthetic microstructures {include} Neper \cite{quey2011large} and Dream3d \cite{groeber2014dream}. They generate synthetic granular microstructures with given characteristics such as grain size, grain sphericity, and grain aspect ratio based on Voronoi tessellation \cite{schrefl1992numerical}. The grain structure has to be modified further, in order to include grain boundary phases. Additional shells around the grains with modified intrinsic magnetic properties may be required in order to represent soft magnetic defect layers or grain boundary diffusion. These modifications of the grain structure can be achieved using computer aided design tools such as Salome \cite{ribes2007salome}. In particular, boundary phases of a specified thickness can be introduced by moving the grain surfaces by a fixed distance along their surface normal. When the finite element method is used for computing the magnetostatic potential, the magnet has to be embedded {within} an air box. The external mesh is required to treat the boundary conditions at infinity. As a rule of thumb the problem domain surrounding the magnet {should have} at least 10 times the extension of the magnet \cite{chen1997review}. The polyhedral geometry, the grain boundary phase, and the air box that surrounds the magnet are then meshed using a tetrahedral mesh generator. Public domain software packages for mesh generation {include} NETGEN \cite{schoberl1997netgen}, Gmsh \cite{geuzaine2009gmsh}, and TetGen \cite{si2015tetgen}. Ott et al. \cite{ott2009effects} and Fischbacher et al. \cite{fischbacher2017nonlinear} used NETGEN for meshing nanowires with various tip shape and for meshing polyhedral NdFe$_{12}$ based magnets. Zighem et al. \cite{zighem2014magnetization} and Liu et al \cite{liu2017micromagnetic} used Gmsh to mesh complex shaped Co-nanorods and to mesh polyhedral models for Cerium substituted NdFeB magnets, respectively. Fischbacher et al. \cite{fischbacher2017searching} used TetGen to mesh polyhedral core-shell grains separated by a grain boundary phase.

Figure \ref{fig_poly}a shows a synthetic grain structure created with Neper \cite{quey2011large}. The grain size follows a log-normal distribution. The edge length of the cube containing the 27 grains is 300~nm. The thickness of the grain boundary phase is 6~nm. The material parameters for the main phase and the grain boundary phase were the same as used previously. Figure \ref{fig_poly}c and \ref{fig_poly}d show slices through the tetrahedral mesh of the magnet. 

In both cases the mesh size at the boundary is 2.7 nm. In (c) an almost uniform mesh ($a = 1.01$) was created with an average edge length of 2.9 nm and a maximum of 6 nm  in the center of the grains. The number of elements in the magnet is 11.8 millions. In (d) a graded mesh with an expansion factor of 1.09 was created. The average mesh size is then 3.3 nm with a maximum of 11.6 nm. The number of elements is reduced by almost 42 percent to 6.9 million elements. For both meshes we obtain identical demagnetization curves shown in figure \ref{fig_poly}b. For the preconditioned conjugate gradient \cite{exl2018preconditioned} used in this study, the time to solution scales linearly with the problem size. Thus the use of geometric meshes reduces the computation time by a factor of 1/2.  

\section{Rare-earth efficient permanent magnets}

\subsection{Shape enhanced coercivity}

Shape-anisotropy based permanent magnets have a long history. AlNiCo permanent magnets contain elongated particles that form by phase separation during fabrication \cite{becker1968permanent,livingston1981review,jimenez2014advanced}. AlNiCo magnets were usurped as the magnets with the highest energy density product {before} the development of permanent magnets based on high uniaxial magnetocrystalline anisotropy; first ferrite magnets and then those based on rare earths \cite{becker1968permanent}. In fact, some have suggested that the development of iron-rare-earth magnets was initially motivated by the perceived need to replace Co, at that time considered strategic and critical \cite{hadjipanayis1983new}. If true, this  situation ironically mirrors our current plight. Today shape-anisotropy magnets are again sought as candidates for rare-earth free magnets.     

Livingston \cite{livingston1981review} and Ke et al. \cite{ke2017simulation} discussed the coercivity mechanism{s} of shape-anisotropy based permanent magnets. In ellipsoidal particles the demagnetizing field is uniform. If the particles are small enough to reverse by uniform rotation \cite{frei1957critical} the change of the demagnetizing field with the orientation of the magnet leads to an effective uniaxial anisotropy $K_\mathrm{d} = (\mu_0/2)M_\mathrm{s}^2 (N_\perp-N_\parallel)$ \cite{kronmuller1987angular}, whereby $N_\perp$ and $N_\parallel$ are the demagnetizing factors perpendicular and normal to the long axis of the ellipsoid. However, if the particle diameter becomes too large magnetization reversal will be non-uniform and coercivity drops \cite{livingston1981review}. Coercivity {also decreases} with {increased} packing density of the particles \cite{kittel1949physical}.  

\begin{figure}
\includegraphics[width=8.0cm]{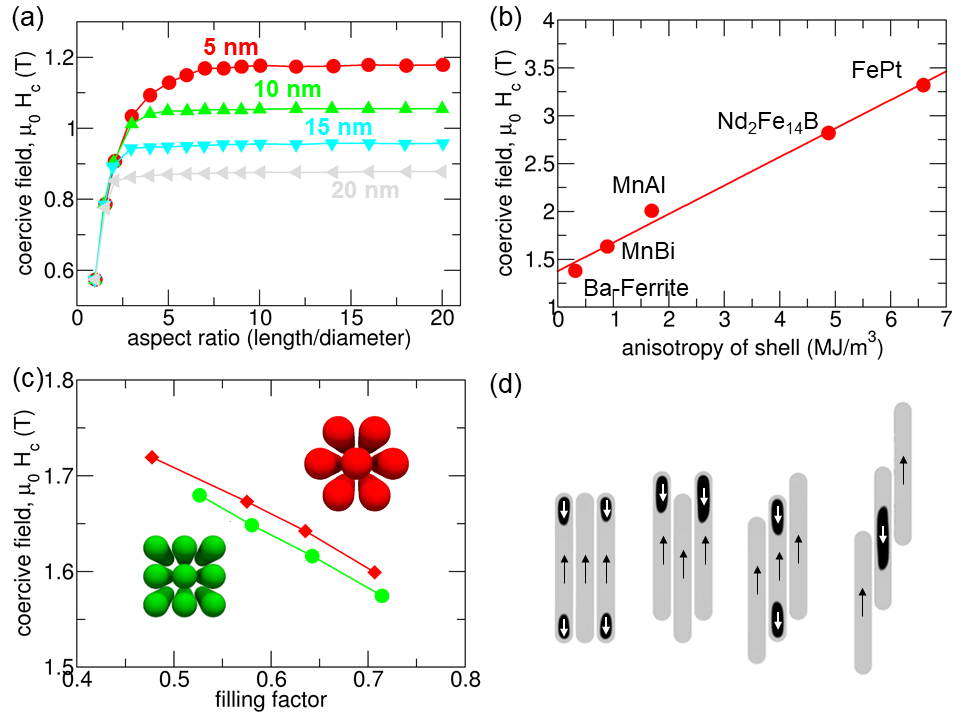}
\caption{\label{fig_Co} (a) Coercive field as function of the aspect ratio of a Co nanorod for different diameters. (b) Coercive field of a Co wire with an aspect ratio of 10:1 and a diameter of 10 nm as a function of the magnetocrystalline anisotropy of a hard magnetic shell. (c) Reduction of the nucleation field of FePt-coated Co nanorods as function of packing density. (d) Formation of reversed domains in three interacting FePt-coated Co nanorods depending on their relative position.}
\end{figure}

\subsubsection{Magnetic nanowires}

High aspect ratio Co, Fe, or CoFe nanowires can be grown via a chemical nanosynthesis polyol process or electrodeposition \cite{maurer2007magnetic,niarchos2015toward,palmero2016synthesis,ener2017consolidation}. Key microstructural features of nanowires and nanowire arrays such as particle shape \cite{ott2009effects}, packing density and alignment \cite{bance2014micromagnetics,toson2015nanostructured,ke2017simulation}, and particle coating \cite{toson2015nanostructured} have been studied using micromagnetic simulations. 

The shape of the ends of magnetic nanowires affects the coercivity. An improvement in the coercive field of between 5 and 10 percent is found when the ends are rounded, as opposed to being flattened like an ideal cylinder \cite{maurer2007magnetic,bance2014micromagnetics}. This enhancement of the coercive field is due to the reduction of high demagnetizing fields which occur at the front plane of the cylinder \cite{holz1968theoretische}. One of the important results from the shape anisotropy work is that the width of elongated nanoparticles is more crucial than the length. Assuming that a particular aspect ratio of 5:1 has been reached, increasing the length will give no further increase in coercive field. Figure \ref{fig_Co}a gives the computed coercive field of Co cylinders ($K = 0.45$~MJ/m$^3$, $\mu_0 M_\mathrm{s} = 1.76$~T, $A = 1.3$~J/m) with rounded ends as function of the aspect ratio for different cylinder diameters. The smaller the diameter the higher is the coercive field. {Ener et al. \cite{ener2017consolidation} measured the coercive field of Co-nanorods for diameters of 28~nm, 20~nm, and 11~nm to be 0.36~T/$\mu_0$, 0.47~T/$\mu_0$, 0.61~T/$\mu_0$, respectively. When comparing with micromagnetic simulations we have to consider misorientation, magnetostatic interactions, and thermal activation which occur in the sample but are not taken into account in the results presented in figure \ref{fig_Co}a. Viau et al. \cite{viau2009highly} measured a coercive field of 0.9~T/$\mu_0$ at $T = 140$~K for Co wires with a diameter of 12.5~nm.} 

\subsubsection{Nanowires with core-shell structure}

The coercivity of Fe nanorods can be improved by adding anti-ferromagnetic capping layers at the end. Toson et al. \cite{toson2015nanostructured} showed that exchange bias between the  antiferromagnet caps and the Fe rods mitigates the effect of the strong demagnetizing fields and thus increases the coercive field by up to 25 percent. Alternatively, a Co cylinder may be coated with a hard magnetic material. Figure \ref{fig_Co}b shows the coercive fields of a Co-nanorod with a diameter of 10 nm and an aspect ratio of 10:1 which are coated by a 1 nm thick hard magnetic phase. The coercive field increases linearly with the magnetocrystalline anisotropy constant of the shell.

The demagnetizing field of one rod reduces the switching field of another rod close-by. The closer the rods, the stronger is this effect. We simulated two bulk magnets consisting of either a hexagonal close-packed (h.c.p.) or a regular 3 x 3 arrangement of Co/FePt core-shell rods. As the filling factor increases, the separation of the nanorods becomes smaller, meaning that the demagnetizing effects on neighboring rods increase, so the nucleation field leading to reversal is reduced (see figure \ref{fig_Co}c). Depending on the arrangement of the nanorods, the magnetostatic interaction field {nucleates reversal} in neighboring nanowires (see figure \ref{fig_Co}d). The reversed {regions} start to grow in the core of the wire owing to the hard magnetic shell.

\subsection{Grain boundary engineering}

There is evidence from both micromagnetic simulations {\cite{bance2014influence,bance2015thermally,helbig2017experimental}} and experiments \cite{helbig2017experimental} that magnetization reversal in conventional magnets starts from the surface of the magnet or the grain boundary. An obvious cure to improve the coercivity of NdFeB magnets is local enhancement of the anisotropy field near the grain surface {\cite{bance2015thermally}}. This may be achieved by adding heavy rare-earth elements such as Dy in a way that (Dy,Nd)$_2$Fe$_{14}$B forms only near the grain boundaries, creating a hard shell-like layer. Possible routes for the latter process are the addition of Dy$_2$O$_3$ as a sintering element \cite{ghandehari1987microstructural} or by grain boundary diffusion  \cite{hirota2006coercivity,sodervznik2016high}. These production techniques reduce the share of heavy rare-earth elements while maintaining the high coercive field of (Dy,Nd)$_2$Fe$_{14}$B magnets. In addition, grain boundary diffused magnets show a higher remanence, because the volume fraction of the (Dy,Nd)$_2$Fe$_{14}$B phase, which has a lower magnetization than Nd$_2$Fe$_{14}$B, is small. Similarly, the coercive field has been enhanced by Nd-Cu grain boundary diffusion which reduced the Fe content in the grain boundaries \cite{sepehri2013high}.

\begin{figure}
\includegraphics[width=8.0cm]{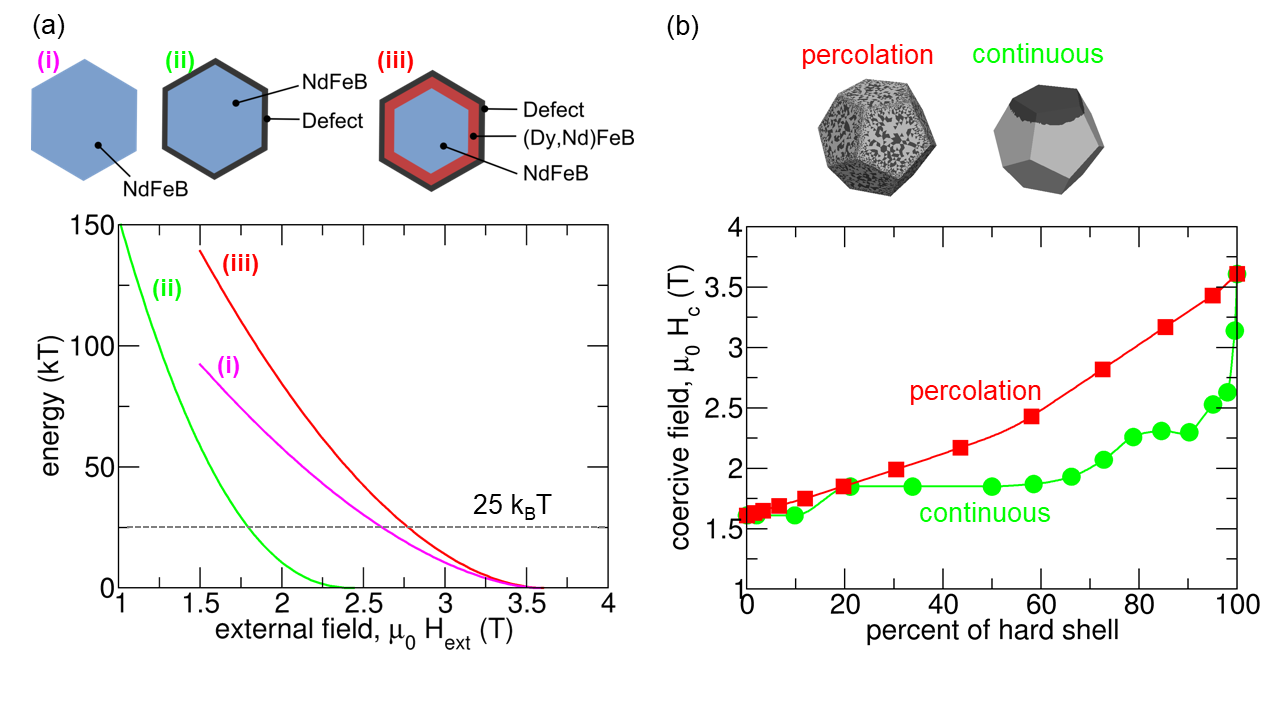}
\caption{\label{fig_coreshell} (a) Energy barrier for magnetization reversal as function of the applied field for (i) a perfect Nd$_2$Fe$_{14}$B grain, (ii) a Nd$_2$Fe$_{14}$B grain with a surface defect with zero anisotropy, and (iii) a system with a defect and a (Dy$_{47}$Nd$_{53}$)$_2$Fe$_{14}$B shell. The critical field value at which the energy barrier becomes $25k_\mathrm{B}$ is the temperature dependent coercive field. $T = 450$~K. {Data taken from \cite{bance2015thermal}}. (b) Coercivity of a Nd$_2$Fe$_{14}$B particle as function of the percentage of coverage with a Tb-containing shell for the continuous coverage model and the percolation model.}
\end{figure}

\subsubsection{Core-shell grains}

We used the string method \cite{weinan2007simplified,bance2015thermally} to compute the temperature-dependent hysteresis properties of Nd$_2$Fe$_{14}$B permanent magnets in order to assess the influence of a soft outer defect and a hard shell created by Dy diffusion. Dodecahedral grain models, approximating the polyhedral geometries of grains observed in actual rare earth permanent magnets, are prepared in three varieties: (i) a pure NdFeB ($K=2.1$~MJ/m$^3$, $\mu_0 M_\mathrm{s} = 1.3$~T, $A = 4.9$~pJ/m) grain with no defect and no shell, (ii) a NdFeB core with a soft outer defect ($K=0$) of 2~nm thickness and (iii) a Nd$_2$Fe$_{14}$B  core with a (Dy$_{47}$Nd$_{53}$)$_2$Fe$_{14}$B hard shell ($K=2.7$~MJ/m$^3$, $\mu_0 M_\mathrm{s} = 1$~T, $A = 6.4$~pJ/m) of 4~nm plus an outer defect (2nm). The outer grain diameter is constant at 50~nm. Figure \ref{fig_coreshell}a shows how the energy barrier decreases as a function of applied field. In all model variations, at $T = 450$~K the thermal activation reduces the coercivity by around 25 percent. The reduction in coercivity from the soft defect in (ii) is canceled out by the hard shell in (iii).  

In a real magnet the diffusion shell will not necessarily be of uniform thickness or fully cover the grain. We {investigate} this effect for {a Nd$_2$Fe$_{14}$B grain} with a diameter of 250~nm and {a} Tb-containing shell. In order to investigate the effects of imperfect shells we {simulate} systems where parts of the material in a 20 nm thick shell {are} replaced with the core material, in order to calculate the change in coercive field. A number of approaches are possible. First, a continuous island of varying size is formed where the (Tb$_{0.5}$Nd$_{0.5}$)$_2$Fe$_{14}$B is replaced by Nd$_2$Fe$_{14}$B. A 2 nm outer defect layer is still present, with material properties of elements matching those of the material they cover, except that $K = 0$. A second approach for an imperfect hard magnetic shell is percolation. Random shell elements {are} switched to the core material. In the beginning the islands with the core materials {are} very small until the number of switched elements {increase} and the islands {join} up. Depending on the type of the Tb-containing shell the behavior is different. The coercive field is plotted against the percentage of Tb-containing material in the shell in figure \ref{fig_coreshell}b for continuous coverage and {for percolated coverage}.  For the continuous  coverage model there is an exponential relationship, with a more complete covering leading to the highest coercivity values. As soon as the covering is reduced, the coercivity drops rapidly. The trend is not smooth since at various points the growing island's boundary reaches the edges and corners of the dodecahedral grain, locations of importance where reversal begins. For the percolation model the coercive field increases linearly with the amount of (Tb,Nd)$_2$Fe$_{14}$B in the shell.

\begin{figure}
\includegraphics[width=8.0cm]{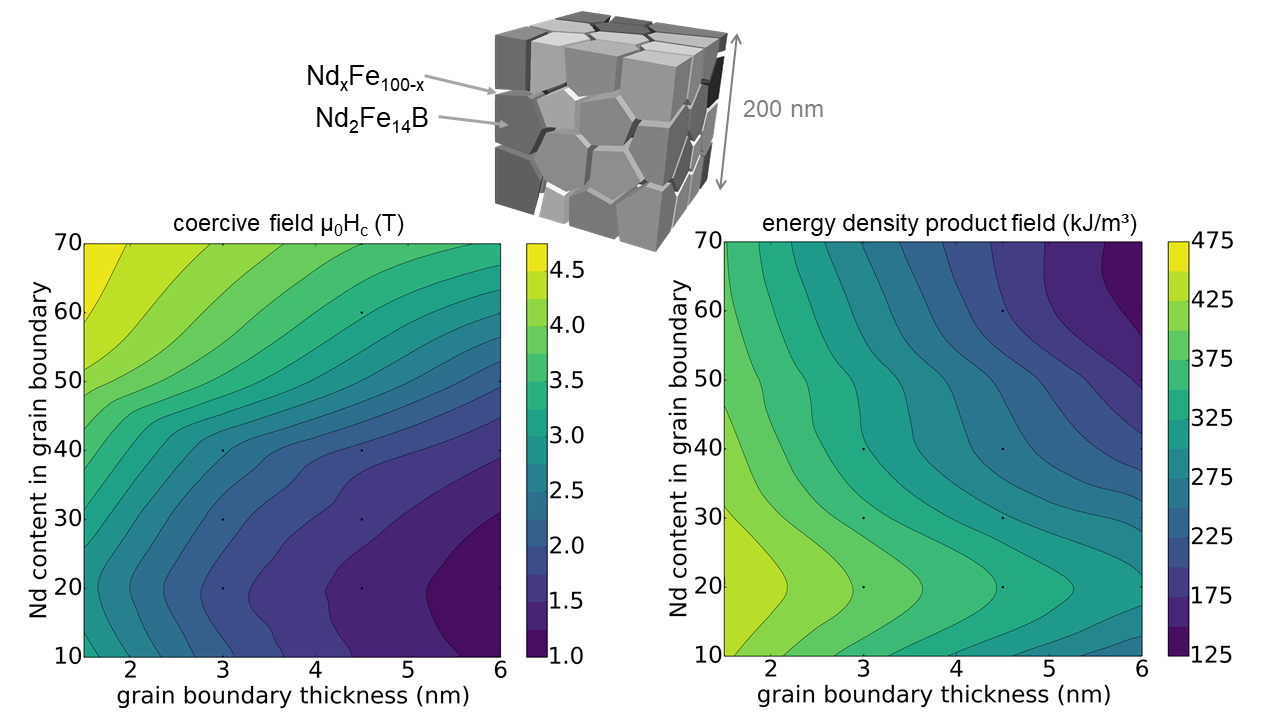}
\caption{\label{fig_gb} Coercive field (left) and energy density product (right) as function of the grain boundary properties for a nanocrystalline magnet with a grain size of 50~nm. The computed demagnetization curves are corrected with a demagnetizing factor of $N=1/3$. The coercive field is given with respect to the internal field $H$.}
\end{figure}

\subsubsection{Grain boundary properties}

By NdCu diffusion high performance Nd$_2$Fe$_{14}$B magnets without any heavy rare earths can be achieved \cite{akiya2014high,liu2016coercivity}. Energy-dispersive X-ray spectroscopy and atom probe analysis \cite{sepehri2013high} showed the formation of a Nd-rich intergranular phase upon infiltration. The Nd rich grain boundary phase predominantly forms at the grain surfaces perpendicular to the anisotropy axes \cite{sepehri2013high}. {The coercive field of hot deformed NdFeB magnets increases from 1.5~T/$\mu_0$ to 2.3~T/$\mu_0$ by Nd-Cu infiltration. Infiltration increases the Nd concentration from 38 at.\% to 80 at.\% in grain boundary perpendicular to the anisotropy axes, whereas the Nd content in grain boundaries parallel to the anisotropy axes remains low reaching only 25 at.\% after infiltration.} 

In order to understand the influence of the grain boundary properties on coercivity and energy density product we {compute} demagnetization curves of a nanocrystalline magnet. We {vary} the thickness of the grain boundary from 1.5~nm to 6~nm, keeping the size of the magnet constant. We also {change} the Nd content of the grain boundary phase and {adjust} the magnetization and the exchange constant of the grain boundary phase according to the data published by Sakuma et al. \cite{sakuma2015magnetism}. Please note that the magnetization of Nd$_x$Fe$_{100-x}$ shows a maximum at $x = 20$. We correct the demagnetization curve with the macroscopic demagnetization factor $N=1/3$ and extract the coercive field and the energy density product (see figure \ref{fig_gb}). Clearly the maximum coercive field is reached for a thin Nd-rich grain boundary phase. For nanocrystalline grains the magnetization of the grain boundary phase contributes to the total magnetization. Therefore, the maximum energy density product occurs for a Nd content of 20 percent and a grain boundary thickness of 1.5~nm. A similar result was reported by Lee et al. \cite{lee2017magnetization} who simulated the hysteresis properties as {a} function of the magnetization and the exchange constant in the grain boundary phase.    

\subsection{Alternative hard magnetic compounds}

In the following we describe how coercive field, remanence, and energy density product change with typical microstructural features for several possible alternative hard magnetic phases. 

For all simulations we {assume} aligned grains. The alignment factor $f = \cos(\phi)$ {is} always close to unity. Here $\phi$ is the average misalignment angle. To account for higher misalignment the $M_\mathrm{r}$-values and the $(BH)_\mathrm{max}$-values need to be multiplied with $f$ and $f^2$, respectively.  Let us consider the following example: The simulated values are $\mu_0 M_\mathrm{r} = 1.22$~T and $(BH)_\mathrm{max} = 251$~kJ/m$^3$ for a L1$_0$ FeNi magnet. Assuming $\phi = 20$ degrees the expected values for the remanence and the energy density products are $\mu_0 M_\mathrm{r} = 1.15$~T and $BH_\mathrm{max} = 222$~kJ/m$^3$.

\begin{figure}
\includegraphics[width=8.0cm]{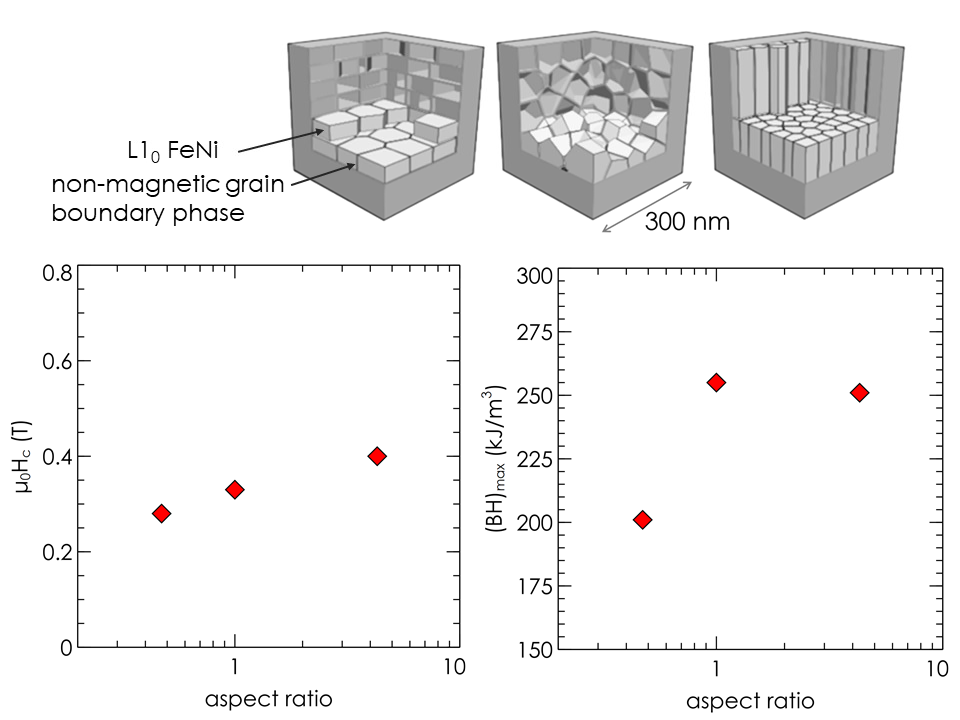}
\caption{\label{fig_feni} Magnetic properties as function of the aspect ratio of the grains in a L1$_0$-FeNi granular system with $K = 0.35$~MJ/m$^3$. Top: Nanostructures with different aspect ratios of the grains. Left: Coercivity. Right: Energy density product. The computed demagnetization curves are corrected with a demagnetizing factor of $N=1/3$. The coercive field is given with respect to the internal field $H$. {Data taken from \cite{kovacs2017micromagnetic}}.}
\end{figure}

\subsubsection{L1$_0$-FeNi based permanent magnets}

The rare-earth free FeNi with a tetragonal L1$_0$ structure has a large saturation magnetization of $\mu_0 M_\mathrm{s} = 1.5$~T{,} which translates to a theoretically possible energy product of $(BH)_\mathrm{max} = 448$~kJ/m$^3$. However, such a high energy density product requires a sufficiently large magnetocrystalline anisotropy. The empirical law $\kappa > 1$ suggests an anisotropy constant $K > 1.8$~MJ/m$^3$. Lewis and co-workers \cite{lewis2014inspired,poirier2015intrinsic} studied the crystal lattice, microstructure and magnetic properties of the meteorite NWA 6259. Its L1$_0$-FeNi phase is highly ordered and therefore regarded as a possible candidate for {use in} permanent magnets. They estimated the magneto-crystalline anisotropy of the meteorite to be $K = 0.84$~MJ/m$^3$. Edstr\"om et al. \cite{edstrom2014electronic} predicted an anisotropy in the range of 0.48~MJ/m$^3$ to 0.77~MJ/m$^3$, using density functional theory. The magnetocrystalline anisotropy linearly depends on the chemical order parameter \cite{kojima2011magnetic}. 

{Whereas chemical ordering is much smaller in most other attempts to fabricate L1$_0$ FeNi \cite{takanashi2017fabrication}, Goto et al. \cite{goto2017synthesis} synthesized L1$_0$FeNi powder with a degree of order of 0.7 through nitrogen insertion and topotactic extraction. They measured a coercive field of 0.18~T/$\mu_0$.}

We investigated how nanostructuring may help to create reasonable hard magnetic properties with a low-anisotropy L1$_0$-FeNi phase.  In L1$_0$-FeNi thin films made by combinatorial sputtering an anisotropy constant $K = 0.35$~MJ/m$^3$ was measured by ferromagnetic resonance \cite{kovacs2017micromagnetic}. We computed the demagnetization curves for three different nanostructures consisting of platelets, equiaxed grains, and columnar grains. The grains have approximately the same volume of $72 \times 72 \times 34$~nm$^3$, $56 \times 56 \times 56$~nm$^3$, and  $34 \times 34 \times 146$~nm$^3$, for the platelets, polyhedra, and columns, respectively. The macroscopic shape of the magnet is cubical with an edge length of 300~nm. The volume fraction of {the} non-magnetic grain boundary phase is 18 percent.

The coercive field increases with increasing aspect ratio. The data summarized in figure \ref{fig_feni} shows that the coercivity can be tuned by 120~mT/$\mu_0$ through a change {in} the shape of the grains. The grain shape has no influence on the remanent magnetization which was computed to be $\mu_0 M_\mathrm{r} = 1.21$~T. For platelet-shaped grains the energy density product is coercivity limited with $(BH)_\mathrm{max} = 201$~kJ/m$^3$. For higher coercivity such as in equiaxed and columnar grains the expected energy density product is close to $(BH)_\mathrm{max} = 255$~kJ/m$^3$. 
 
\begin{figure}
\includegraphics[width=8.0cm]{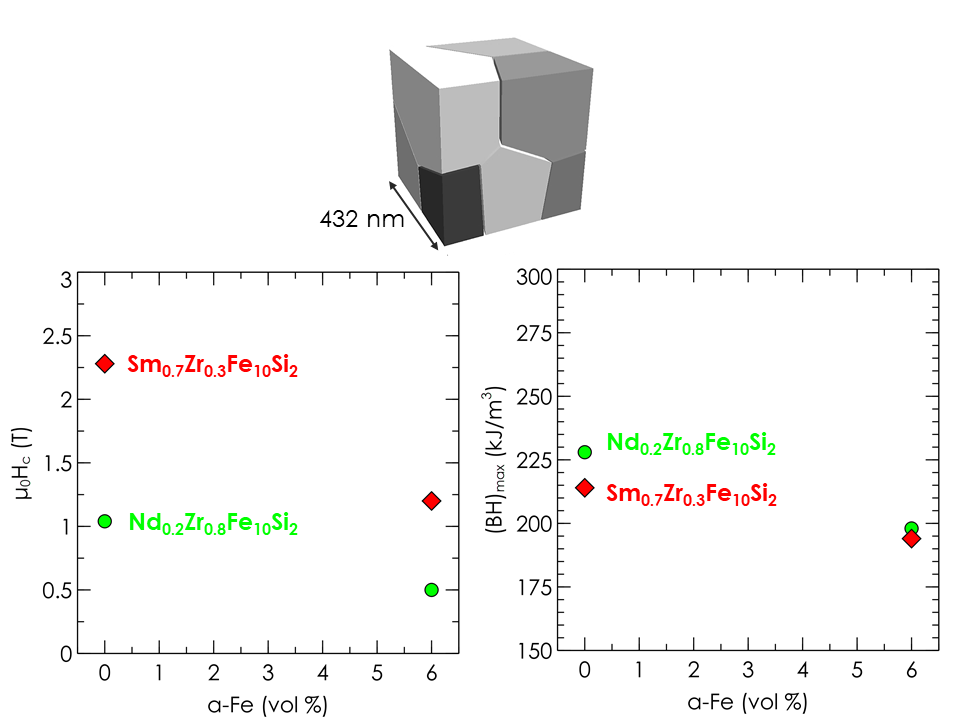}
\caption{\label{fig_ThMn12} Magnetic properties of Nd$_{0.2}$Zr$_{0.8}$Fe$_{10}$Si$_2$ and Sm$_{0.7}$Zr$_{0.3}$Fe$_{10}$Si$_2$ without and with $\alpha$-Fe inclusions.Top: Granular structure used for the simulation. Left: Coercivity. Right: Energy density product. The computed demagnetization curves are corrected with a demagnetizing factor of $N=1/3$. The coercive field is given with respect to the internal field $H$.}
\end{figure} 
 
\subsubsection{ThMn$_{12}$ based permanent magnets}

Possible candidate phases are NdFe or SmFe compounds in the ThMn$_{12}$ structure, which {were} discussed already in the late 1980s \cite{buschow1988structure,hu1989intrinsic}. However, NdFe$_{12}$ and SmFe$_{12}$ are not stable without any stabilizing elements such as Ti, Mo, Si, or V \cite{de1988some,ohashi1988magnetic}. At 450~K Nd and Sm based magnetic phases in the ThMn$_{12}$ structure show a higher magnetization and a higher anisotropy field than Nd$_{2}$Fe$_{14}$B \cite{hirayama2015ndfe,kuno2016sm}.  In addition, the rare earth to transition metal ratio of the 1:12 based magnets is lower. Therefore, magnets based on this phase are considered as a possible alternative to Nd$_2$Fe$_{14}$B magnets \cite{hirosawa2017perspectives,gabay2017recent}. The rare-earth content is further reduced if some Nd or Sm is partially replaced with Zr \cite{gabay2017recent}. {The fabrication of a magnets in the 1:12 structure is difficult. In contrast to Nd$_2$Fe$_{14}$B, phases in the vicinity of R(Fe,M)$_{12}$ (R rare earth, M stabilizing element) in the equilibrium phase diagram are ferromagnetic. As a consequence there is no isolation of the grains with a non-magnetic or only weakly ferromagnetic grain boundary phase \cite{gabay2017recent}. Gabay and Hadjipanayis \cite{gabay2017mechanochemical} measured a coercive field of 1.08~T/$\mu_0$ in Sm$_{0.3}$Ce$_{0.3}$Zr$_{0.4}$Fe$_{10}$Si$_2$ oriented particles prepared by a mechano-chemical route.}  

Here we look at the potential of the very rare-earth lean compounds Nd$_{0.2}$Zr$_{0.8}$Fe$_{10}$Si$_2$ \cite{gjoka2016effect} and Sm$_{0.7}$Zr$_{0.3}$Fe$_{10}$Si$_2$ \cite{gabay2016structure}. Experiments show that the magnets contain $\alpha$-Fe as a secondary phase with a volume fraction of about 6 percent \cite{gjoka2016effect}. Therefore, we {investigate} the influence of the $\alpha$-Fe content on the hysteresis properties. The synthetic microstructure used for the simulations is shown in figure \ref{fig_ThMn12}. The volume fraction of the grain boundary phase is 8 percent. The grain boundary phase was assumed to be moderately ferromagnetic with $\mu_0 M_\mathrm{s} = 0.56$~T and $A = 2.5$~pJ/m. Nd$_{0.2}$Zr$_{0.8}$Fe$_{10}$Si$_2$ shows uniaxial anisotropy \cite{gjoka2016effect} with an anisotropy constant of $K = 1.16$~MJ/m$^3$ and magnetization of $\mu_0 M_\mathrm{s} = 1.12$~T \cite{gjoka2016effect}. For Sm$_{0.7}$Zr$_{0.3}$Fe$_{10}$Si$_2$ we {use} $K = 3.5$~MJ/m$^3$ and $\mu_0 M_\mathrm{s} = 1.08$~T \cite{gabay2016structure}. We {compare} two scenarios: (i) A sample without any $\alpha$-Fe as secondary phase, and (ii) a sample in which each grain contains an $\alpha$-Fe inclusion so that the total volume fraction of $\alpha$-Fe is 6 percent.

The presence of $\alpha$-Fe reduces the coercive field. In Nd$_{0.2}$Zr$_{0.8}$Fe$_{10}$Si$_2$ it decreases by about a factor of 1/2 from $\mu_0 H_\mathrm{c} = 1.04$~T to $\mu_0 H_\mathrm{c} = 0.5$~T. Similarly, in Sm$_{0.7}$Zr$_{0.3}$Fe$_{10}$Si$_2$ the coercive field changes from $\mu_0 H_\mathrm{c} = 2.28$~T to $\mu_0 H_\mathrm{c} = 1.2$~T when $\alpha$-Fe inclusions are taken into account. The remanent magnetization for the Nd and Sm compound {is} $\mu_0 M_\mathrm{r} = 1.07$~T and $\mu_0 M_\mathrm{r} = 1.04$~T, respectively.  The presence of $\alpha$-Fe reduces the remanent magnetization by 4 percent and 3 percent in the Nd and the Sm magnet, respectively. With $\alpha$-Fe the energy density product reduces from 228~kJ/m$^3$ to 198~kJ/m$^3$ and from 214~kJ/m$^3$ to 194~kJ/m$^3$ in Nd$_{0.2}$Zr$_{0.8}$Fe$_{10}$Si$_2$ and Sm$_{0.7}$Zr$_{0.3}$Fe$_{10}$Si$_2$, respectively.

\section{Summary}

Hard magnetic phases for rare-earth free or rare-earth reduced permanent magnets may show a lower magnetocrystalline anisotropy than Nd$_2$Fe$_{14}$B. Therefore a detailed understanding of the influence of the microstructure on the magnetic properties is of utmost importance for the development of new  permanent magnets. Computational micromagnetics reveals the main microstructural effects on the coercive field, the remanence, and the energy density product.

\subsection{Grain boundary phase} The grain boundary phase significantly influences the coercive field. If the grain boundary phase is ferromagnetic, the coercive field decreases with increasing thickness of the grain boundary. Dy or Tb diffusion recovers the coercivity of magnets with ferromagnetic grain boundary phases. A heavy rare-earth containing shell with a thickness of 10 nm doubles the coercive field which keeps increasing moderately with further increasing thickness of the hard magnetic shell. High energy products and reasonable coercive fields can be achieved for ferromagnetic grain boundaries {with thicknesses} below 3 nm. In nanocrystalline magnets the remanence and energy density product increase with increasing magnetization of the grain boundary phase.

\subsection{Grain shape} In magnets based on CoFe nanorods coercivity is mostly governed by the thickness of the rods. The highest coercive fields can be obtained if the rod diameter is comparable with the exchange length of the material. Nanostructuring is essential and helps to improve the {hysteresis} loop squareness in materials with low magnetocrystalline anisotropy. For the magnetic properties of commonly synthesized L1$_0$ FeNi the energy product is coercivity limited. A change from platelet-shaped grains to columnar grains may increase the energy density product by 25 percent.

\subsection{Soft magnetic secondary phases} Soft magnetic inclusions may reduce the coercive field by up to a factor of 1/2. If the magnetocrystalline anisotropy is sufficiently high{,} such as in SmFe or NdFe compounds in the ThMn$_{12}$ structure{,} still excellent hard magnetic properties can be achieved despite the presence of $\alpha$-Fe. The reduction of the energy density product by soft magnetic inclusions {is} about 10 percent. 

\section{Conclusion}

{The coercivity and the energy density product were computed for several rare-earth reduced and rare-earth free permanent magnets using micromagnetic simulations. For some materials the theoretically predicted values are higher than those currently achieved in experiments. This discrepancy emphasizes the importance of the microstructure. A small grain size, thin non-magnetic grain boundary phases that separate the grains, and elongated grains for phases with low magnetocrystalline anisotropy are essential to achieve a high coercive field.}

\ack{This work was support by the EU FP7 project ROMEO (Grant no 309729), the EU H2020 project Novamag (Grant no 686056), the Austrian Science Fund FWF (Grant no F4112 SFB ViCoM), the Japan Science and Technology Agency (JST, CREST), Siemens AG, Toyota Motor Corporation, and the future pioneering program “Development of Magnetic Material Technology for High-efficiency Motors” commissioned by the New Energy and Industrial Technology Development Organization (NEDO).
}

\section*{References}
\bibliography{magnets}

\end{document}